\documentclass[11pt]{article}
\usepackage{amsmath}
\usepackage{amssymb}
\usepackage{graphicx}
\usepackage{mathtools}
\usepackage{float}
\begin{document}
%CAMBIATO TITOLO RISPETTO A DRAFT3

\begin{center}
{ \Large \bf Knotted optical fields: Seifert fibrations,\\ braided open books
and topological constraints} 
\end{center}
\vspace{24pt}

\begin{center}
{\sl Annalisa Marzuoli}\\
Dipartimento di Matematica ``Felice Casorati",
Universit\`a  di Pavia\\
via Ferrata 5, 27100 Pavia-I\\
INFN, Sezione di Pavia, via Bassi 6, 27100 Pavia-I \\ 
E-mail: annalisa.marzuoli@unipv.it \\
\vspace{5pt}
{\sl Nicola Sanna}\\
Dipartimento di Fisica ``Giuseppe Occhialini",
Universit\`a  di Milano Bicocca\\
Piazza della Scienza 3, 20126 Milano-I\\ 
E-mail: n.sanna.95@gmail.com 
\end{center}

\vspace{12pt}

\noindent KEYWORDS: Knotted and braided optical fields, framed knots, Seifert fibrations, 
braided open books\\

\noindent {\bf Abstract}\\
This paper presents a novel framework for studying knotted and braided configurations of optical fields, moving beyond the conventional Hopfion solution based on the Hopf fibration. By employing the Seifert fibration, a preferred framing is introduced to characterize the ``knottiness" of tubular neighbourhoods of knots embedded in the 3-sphere. This approach yields a specific presentation of Seifert surfaces and facilitates the description of knotted optical fields, drawing on an enriched version of Ra\~{n}ada's formulation. The relation between knots and braids enables one to explore the connections between configurations generated experimentally through the controlled over- and under-crossings of light beams, and concepts from geometric group theory and algebraic topology. This framework, emphasizing the significance of ``braided open books", sheds light on topological constraints inherent in classes of braids suitable for representing interlaced optical fields. These constraints primarily pertain to braids with $n\leq3$ strands --aligning with existing experimental implementations-- and homogeneous braids with any number of strands. This paper establishes a theoretical foundation for understanding and manipulating knotted optical fields, suggesting potential applications and avenues for further research in this domain.

%%%PROMEMORIA 
%labellare formule dopo \begin{equation}
%per richiamare una formula \eqref{....} -mette automaticamente le parentesi tonde
%citazione bibliografica \cite{label} per singola
%citazione multipla \cite{label1,label2}

\vfill
\newpage
% RIPRESO LAVORO 25-06-24 
%%% VIRGOLETTE APERTE E CHIUSE ``tubular" 

%%%%%%%%%%%%%%%%%%%%%%%%%SECTION 1 
\section{Introduction}
%%%%%%%%%%%%%%%%%%%%%%%
%%%%%%%%%%%%%%%%%%%%%%
Modern technologies able to handle  optical fields with shaped spatial and temporal features have been recently used to design experiments whose outputs are topological configurations such as braided filaments, knots, and links (multi-component knots). 
A partial list of references pertinent to the present work includes theoretical 
\cite{IrBo,Irv,KeBiPe,BoDeFo,ArTiTr}
as well as experimental papers \cite{LaSuMo,SuDrOt},
often supported by numerical simulations.
These configurations have long been considered toy models, and their relevance to modern technologies has yet to be determined. Some attempts have recently been made to use light beams --mathematically ``framed" knots-- as information carriers in the contest of topological quantum computing \cite{LaDeFe}.
\\

The study of such exotic structures bears on a formulation of Maxwell theory in terms of complex scalar fields originally formulated by  Ra\~{n}ada  \cite{Ran89,Ran92} and improved in \cite{RaTr95,RaTr06}.
To set the stage for the substantial extensions proposed in the present work, let us briefly review the Ra\~{n}ada construction and the associated Hopfion solution following Section 2 of the review \cite{ArBoTr} --which we refer to also for exhaustive accounts of the various theoretical
approaches developed over the years.\\
The basic ingredients are two complex scalar fields 
\begin{equation}
\label{ranada1}
	\phi(\mathbf{r}, t), \theta(\mathbf{r},t): \mathbb{R}^3 \times \mathbb{R} 
\rightarrow \mathbb{C}
\end{equation}
associated with the magnetic and electric fields, respectively. This assumption
allows for a direct study of the properties and the evolution of the magnetic and electric lines since they coincide with the level curves of $\phi$ and $\theta$. Both scalars must obey suitable boundary conditions: 
{\bf i)} Since the energy of the em field must be finite at spatial infinity,  the conditions  $\phi=0$ and $\theta=0$ for $|\mathbf{r}|\rightarrow \infty$
must be fulfilled independently of the direction; mathematically this amounts to compactify $\mathbb{R}^3$ to
$\mathbb{R}^3 \cup \{\infty\}$ to get $S^3$ (the unit 3-sphere).  
{\bf ii)} Looking at the inverse images of the maps $\phi,\theta: S^3 \times \mathbb{R} \rightarrow \mathbb{C}$, to ensure that the inverses of the value $\{\infty\} \in \mathbb{C}$ do not depend on either direction, it is necessary to compactify also $\mathbb{C}$ to $\mathbb{C} \cup \{\infty\}$ to get $S^2$ (the unit 2-sphere). Then the updated counterpart of Eq. \eqref{ranada1} reads
\begin{equation}
\label{ranada2}
\phi(\textbf{r}, t), \theta(\textbf{r},t): S^3 \times \mathbb{R} 
\rightarrow S^2
\end{equation}
and geometrically, the spatial part of these maps turns out to be naturally identified with the Hopf map $h:S^3 \rightarrow S^2$. In Appendix A, the features of the Hopf fibration are described in detail, highlighting the construction
of projective coordinates whose role is relevant in the following expressions of the electric and magnetic fields in terms of the Ra\~{n}ada fields 
\begin{align}
\label{ranada3}
\mathbf{B}(\mathbf{r},t)&=\frac{\sqrt{a}}{2 \pi i }\; \frac{\nabla\phi \times \nabla \bar{\phi}}{(1+\bar{\phi}\phi)^2} \,;\nonumber \\
\mathbf{E}(\mathbf{r},t)&=\frac{\sqrt{a}c}{2 \pi i }\; \frac{\nabla \bar{\theta} \times \nabla \theta}{(1+\bar{\theta}\theta)^2}\,.
\end{align}
$a$ is a constant with the dimensions of electric flux and value $a= \hbar c / \epsilon_0$ in SI units (and becomes a real number in natural units).
Since $\nabla\phi$ ($\nabla \theta$) and $\nabla \bar{\phi}$ ($\nabla \bar{\theta}$) are perpendicular to field lines of constant $\phi$ ($\theta$), the magnetic (electric, respectively) field is tangent to lines of constant $\phi$ ($\theta$) for each value of the time variable.
The above radiative fields ($\mathbf{E}\cdot \mathbf{B}=0$) are related, respectively,
to the Faraday 2-form $\mathsf{F}= \tfrac{1}{2} F_{\mu \nu} dx^{\mu} \wedge dx^{\nu}$ and its Hodge dual $^{*}\mathsf{F}= \tfrac{1}{2} ^{*} (F_{\mu \nu}) dx^{\mu} \wedge dx^{\nu}$ which, interestingly, can be obtained geometrically by pulling back with the Hopf map the area 2-form from $S^2$ to $S^3$ \cite{RaTr06}. Without going through further details, the Hopf index 
\cite{Whi} characterizes the homotopy classes 
of the complex fields (and can be related to electric and magnetic helicities). 
Thus, the Hopfion is interpreted as a topological soliton with a non-zero Hopf index. It is not a ``knot''. Instead, the field lines are topologically linked circles filling and warping around the Clifford tori embedded in the ``optical'' sphere $S^3$.\\

In this paper, we go beyond the Hopfion solution by directly addressing true knotted and braided configurations embedded in $S^3$ and endowed with suitable ``framings'' required when dealing with light beams. From the topological viewpoint, this program is achieved in Section 2 by resorting to the Seifert fibration  $f: S^3 \setminus \mathcal{K} \rightarrow S^1$ ($\mathcal{K}$ a knot,  $S^3 \setminus \mathcal{K}$ its complement in $S^3$ and $S^1$ the unit circle). After a review of the notions of fibred and framed knots (Sections 2.1 and 2.2), we identify in Section 2.3 a novel type of framing (called ``preferred'') that gives rise to a particular presentation of the Seifert surface (the inverse image of the Seifert map $f$). The relevant topological invariant is provided here by a specific linking number, related in turn to the number of Dehn twists performed in the tubular neighbourhood of the knot. Then, we show how it is possible to adapt the Ra\~{n}ada's setting to describe knotted optical fields.

Exploiting Alexander's theorem \cite{Ale1} (every knot can be presented  as the closure of a braid) in Section 3 we switch to a more  algebraic setting, starting from the definition of $B_n$, the braid group on $n$ strands (Section 3.1). The motivation is, on the one hand, the possibility of complying with current experimental implementations where light beams over-- and under-cross one another to generate braided configurations. On the other hand, computational and algorithmic questions in geometric group theory and algebraic topology might be addressed effectively once rooted in concrete manipulations of optical configurations.  Such configurations, as already mentioned above, must necessarily be endowed with framings. To do so, we first call into play 
``open books'', topological structures that contain the Seifert fibration as a particular case (Section 3.2). The next step requires us to consider``braided'' open books (Section 3.3), introduced in the seminal papers of L. Rudolph \cite{Rud1,Rud2,Rud3}, and the inherent constraints on the types of configurations that are simultaneously braidable and fibrable. It's been proven that both knots generated as closures of braids with $n \leq 3$ strands, and homogeneous braids on an arbitrary number of strands can be framed (the precise definitions can be recovered from the text). It is interesting to note that these classes,
together with previously known cases whose fibrabilty was derived in purely  topological contexts
\cite{Har,Sta} briefly reported in Section 3.2, include all the knots handled so far in the experiments.

In Section 4 (and in the first part of Appendix B) we discuss some topological and combinatorial notions in knot theory, focussing on the issues related to the findings of the previous sections.
In Section 4.1 the class of torus knots is considered, enhancing  their presentations as homogeneous braids. Given the pivotal role played by Seifert surfaces throughout the text, in Section 4.2 we review the Seifert algorithm and the definition of the genus of a given knot. In Section 4.3 we introduce the Alexander--Conway polynomial, an ambient isotopy invariant that encapsulates combinatorial features of knot diagrams --and that can be computed by means of iterative ``skein relations''-- as well as topological information about the associated Seifert surfaces. In the second part of Appendix B, we collect a few remarks about the classification problem in knot theory and the associated algorithmic questions. Here, for completeness, we also report other polynomial invariants that can be  more effective in distinguishing (inequivalent) knots.\\
In Section 5 we discuss at length possible applications and generalisations of our improved approach to the study of knotted optical fields and related topics.\\
The findings of  the present work --equipped with a self--contained 
treatment of topological and algebraic notions-- have been outlined in the Master Thesis 
\cite{Tesi}.

%\vfill
%\newpage

%%%%%%%%%%%%%%%%SECTION 2  %%%%%%%%%%%%%%%%%%%%%%%%%%%%%%%%%%%%%%%
%%%%%%%%%%%%%%%%%%%%%%%%%%%%%%%%%%%%%%%%%%%%%%%%%%%%%%%%
\section{Fibred and framed knots in the 3-sphere}
%%%%%%%%%%%%%%%%%%%%%%%%%%%%%%%%%%%%%%%%%%%%%%%%%%%%%%%%%
%%%%%%%%%%%%%%%%%%%%%%%%%%%%%%%%%%%%%%%%%%%%%%%%%%%%%%%%
A knot $\mathcal{K}$ is an embedding  of the circle $S^1$ into the Euclidean 3-space $\mathbb{R}^3$ or into the unit 3-sphere $S^3$ in $\mathbb{R}^4$ ($S^3$ can be viewed also as the compactification of 
$\mathbb{R}^3$, $\mathbb{R}^3$ 
$\cup$ $\{\infty\}$\,). A link $\mathcal{L}$ is an embedding of a finite collection of   non--intersecting circles that may be linked and knotted together, $\{L_i\} = \mathcal{L},\, i= 1,2,\dots,N$. In what follows we shall deal with knots (one-component links) for definiteness, and any knot  will be equipped with an orientation induced by the choice of an  orientation on $S^1$.\\
Mathematical knot theory is still an active field of research historically aimed to classify knots types, equivalence classes of knots with respect to ambient isotopy (an ambient isotopy is a continuous, orientation--preserving deformation of the knot strand in the ambient 3-dimensional space). Topological invariants of knots are quantities insensitive to ambient isotopies, and thus are able to detect (at least some of the)  intrinsic topological features of  knotted configurations. The simplest knot invariants are numerical, \textit{e.g.} linking and  crossing numbers, but more sophisticate quantities are Laurent polynomials in one or two formal variables, \textit{cf.} Section 4 and Appendix B.\\
In applications to optical fields crucial properties shared by light beams are, on the one hand,  their tubular structure -- basically a thickening of  the core knotted strand-- and the fact that their natural ambient space is the (optical) 3-sphere $S^3$, on the other. From the topological viewpoint both requirements  can
be  formalised by resorting to the notions of \textit{fibred} and \textit{framed} knots. The former complies with the definition of (Seifert) $S^1$--fibrations in an ambient compact oriented 3-manifolds (which in turn are particular instances of \textit{open book structures}, \textit{cf.} Section 3). The latter, the framing, includes the choice of some additional structures on the tubular neighbourhood of the knot, or, equivalently, on a ribbon cut out from its tubular neighbourhood. Note  that the two notions are often used interchangeably, but for our purposes we keep them distinct to a certain extent.

%%%%%%%%%%%%%%%%%%%%%%%%%%%%%%%%%%%%%%%%%%%%%%%%%%%%%%%%%%%%%%%%%
%%%%%%%%%%%%%%%%%%%%%%%%%%%%%%%%%%%%%%%%%%%%%%%%%%%%%%%%%%%%%%%%
\subsection{Fibred knots from Seifert fibrations in $\mathbf{S^3}$}
In general, a  fibration is a map $\pi$ between (topological) manifolds, that projects a \textit{total space} $M$ onto a \textit{base space} $B$, where each point $b$ in the base has a neighbourhood $U$, such that the map  $\pi^{-1}(U) \rightarrow U\times F$
is a homeomorphism (local trivialisation) and  the \textit{fibre} on $b\in B$ is $F_b\,:=\pi^{-1}(b)$, where $F$ is called the \textit{typical fibre}.\\
The Hopf fibration has a paradigmatic role  in the theory of fibre bundles and of compact Lie groups as well. Its rich structure   is encoded into the sequence $S^1 \hookrightarrow S^3 \rightarrow S^2$, involving unit spheres in one, two and three dimensions. The sphere $S^3$ is the total space, $S^2$ the base and $S^1$ the typical fibre, and one says that $S^3$ ``fibres in circles'' over $S^2$, \textit{cf.} Appendix A for a comprehensive review. 

Crucial differences between the Hopf fibration and the $S^1$--fibrations considered from now on \cite{SeTh,Mon,BaGr}, are given by the changes of the total space --still a compact oriented 3-manifold-- and of the base space, here the circle $S^1$ parametrised by $\theta \in [0,2\pi]$.  
An oriented knot $\mathcal{K}\hookrightarrow S^3$ is said to be a \textbf{fibred knot}  if there exists an
$S^1$--fibration map, or Seifert fibration,
	\begin{equation}
	\label{fibra1}
		f\,:\,S^3\backslash \mathcal{K}\; \rightarrow \;S^1\,,
	\end{equation}
where $S^3\backslash \mathcal{K}$ is the complement of $\mathcal{K}$ in the ambient space $S^3$.
%\end{definition}
\noindent The collection  of all  the fibres $\Sigma_\theta:=f^{-1}(\theta)$,   $\theta \in [0,2\pi]$,  gives rise to an oriented compact surface $\Sigma$ in $S^3$ whose  boundary $\partial \Sigma$ is the given knot, and is  referred to as a \textit{Seifert surface} for $\mathcal{K}$. Note that for a given knot (link) there exist more than one Seifert surface, depending on how the knot strand is bent in the ambient space, or on the different choices of planar projections of the knot (more on this issue will be addressed in the following sections).
\begin{figure}[H]
	\includegraphics[width=5cm]{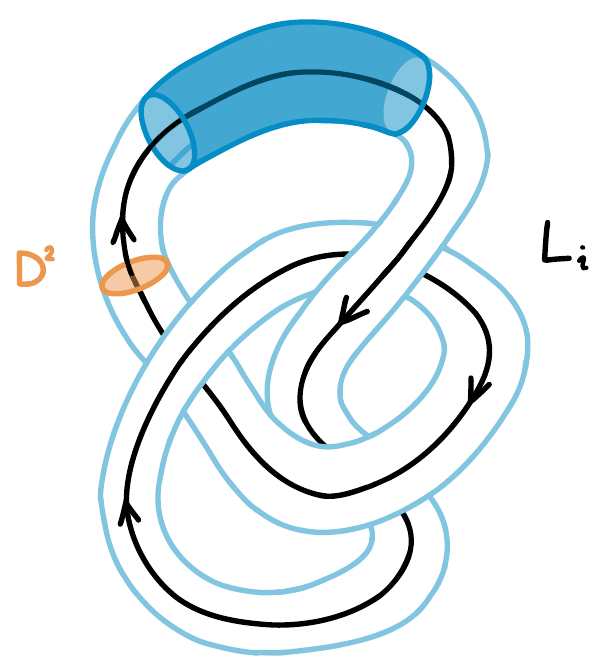}
	\centering
	\caption{The tubular neighbourhood of a component $L_i$ of some link $\mathcal{L}$.}
	\label{tubb}
\end{figure}
The meaning of  \eqref{fibra1} is disclosed by recognizing the inherent structure of a normal $D^2$-bundle over $\mathcal{K} \hookrightarrow S^3$, actually a global trivialisation of the fibre bundle structure.
Roughly speaking, the fibration must be  well behaved near the knot, namely 
the knot must be equipped with a  \textit{tubular neighbourhood} $\mathcal{N}$, 
as depicted in Fig.\ref{tubb}. Parametrising  the  close unit disk as $D^2=\left\lbrace(x,y)\in \mathbb{R}^2: x^2+y^2\leq 1 \right\rbrace$,  
with $\mathcal{K}\cong S^1 \times \left\lbrace 0\right\rbrace$, the  trivialisation property reads  
\begin{equation}
	\label{trivial}
	S^1 \times D^2  \;\cong \; \mathcal{K}  \times D^2\,:= \mathcal{N}\, \hookrightarrow \,S^3 \;,
	\end{equation}
where the  homeomorphism  $\cong$ is by no means unique and its relevance is discussed in the next section. 

\begin{figure}[H]
	\includegraphics[width=8cm]{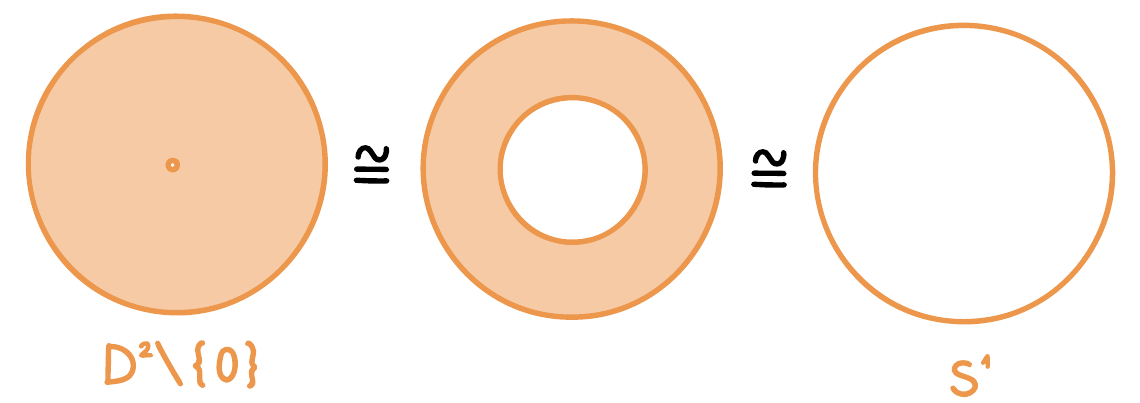}
	\centering
	\caption{The 2-disk with its origin removed --the point where the knot curve goes through-- is continuously deformed to its boundary circle.}
	\label{contract}
\end{figure}

 Then the restriction of the fibration map in \eqref{fibra1} to the tubular neighbourhood with the knot removed  projects onto the second factor, namely on the  $S^1$ which bounds the disk, \textit{cf.} Fig.\ref{contract}, and  represents the base of the Seifert fibration
\begin{align}
\label{fibra2}
	f \;\vert_{S^1\times (D^2\backslash \left\lbrace 0\right\rbrace)}:\;\;\; S^1\times (D^2\backslash \left\lbrace 0\right\rbrace)\;\;\;&\rightarrow \;  \;\; S^1\;\;\; \nonumber \\
	\scriptstyle(x,y) \in \,D^2 \;\;\;\;\scriptstyle&\mapsto \;\;\;\:\scriptstyle\frac{y}{\lvert y \lvert}\,.
\end{align}
It is well--known since the seminal work of Harer \cite{Har} that the  answer to the question ``can all knots (links)  be fibred?'' is  ``no''. In Sections 3.2 and 3.3 we shall address this issue in details.

%%%%%%%%%%%%%%%%%%%%%%%%%%%%%%%%%%%%%%%%%%%%%%%%%%%
%%%%%%%%%%%%%%%%%%%%%%%%%%%%%%%%%%%%%%%%%%%%%%%%%
\subsection{Framed knots and solid tori}

\begin{figure}[H]
	\includegraphics[width=7cm]{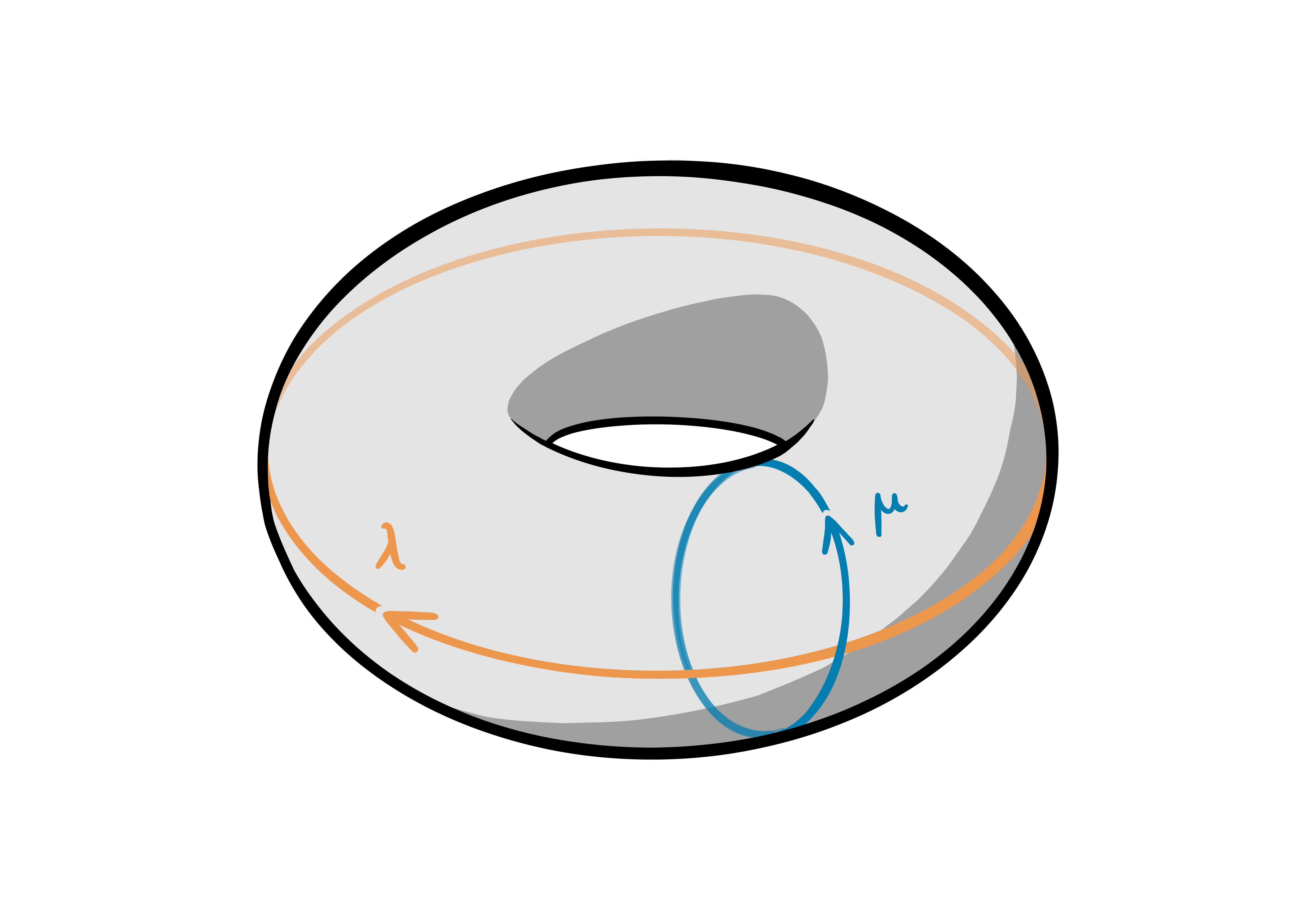}
	\centering
	\caption{The  longitude and meridian on the standard torus.}
	\label{Torus}
\end{figure}

Let $\mathcal{T}$ denotes the solid torus $S^1 \times D^2$ standardly embedded in $\mathbb{R}^3$ or $S^3$. A  \textit{framing} 
of the tubular neighbourhood $\mathcal{N} = \mathcal{K} \times D^2$  is the choice of a specific homeomorphism
\begin{equation}
	\label{framing1}
	\hat{\Phi}\,: \,\mathcal{T} \,:=\, S^1 \times D^2  \;\longrightarrow \; \mathcal{K} \times D^2 \,=\, \mathcal{N} \hookrightarrow S^3.
	\end{equation}
A \textit{meridian} of $\mathcal{N}$ is a simple closed curve $\boldsymbol{\mu}$, lying on the boundary $\partial \mathcal{N}$, which is contractible in $\mathcal{N}$ (but not in the 2-dimensional torus $\partial \mathcal{N}$). A \textit{longitude} of $\mathcal{N}$ is a closed curve $\boldsymbol{\lambda}$ in $\partial \mathcal{N}$ intersecting any  meridian in a single point. The representative meridian and longitude of the standard solid torus $\mathcal{T}$ are depicted in Fig.\ref{Torus}, while meridians and longitudes in $\mathcal{N}$ are their images under the framing map  $\hat{\Phi}$. In both spaces any pair of meridians are ambient isotopic to each other, while there is an infinite number of ambient isotopy classes of longitudes. The choice of a specific class of longitudes will be crucial in the selection of a preferred framing, as will be explained in Section 2.3.
Given $\mathcal{N}$ and a specific $\hat{\Phi}$, the curve $\hat{\Phi} (S^1 \times \{0\})$ can be freely deformed as to coincide with the knot $\mathcal{K}$ itself,   and the choice of a longitude $\hat{\Phi}(S^1 \times \{1\})$ defines a framing of $\mathcal{N}$.
Denoting $\mathcal{K}_{\digamma}$ such a longitude, this construction generates an oriented, knotted and twisted band inside $S^3$ which is a \textbf{framed knot}. Vice-versa, if we equip $\mathcal{K} \hookrightarrow S^3$ with a framing --thickening the knot in the transverse direction as to obtain another curve $\mathcal{K}_{\digamma}$ with the same orientation-- the framing of the whole $\mathcal{N}$ is unique up to ambient isotopy.\\
The description of framed knots as bands (or ribbons) makes it easier to look at their  geometric and topological features. 
The twist $Tw$ of  the band with boundary $(\mathcal{K}, \, \mathcal{K}_{\digamma})$ is simply the number of its twists. The writhe $Wr$ of a planar diagram of  $\mathcal{K}$ (namely the projection of the knot curve onto a plane) is  the total number of positive crossings minus the total number of negative crossings (\textit{cf.} Appendix B). 
These two quantities are combined as to provide a topological invariant (\textit{i.e.} 
an ambient isotopy invariant) \cite{Car}
\begin{equation}
	\label{framing2}
 Lk\, (\mathcal{K}, \mathcal{K}_{\digamma}) \,=\,  Wr( \mathcal{K})\,+\,Tw 
 (\mathcal{K}, \, \mathcal{K}_{\digamma})\,,
\end{equation}
where $Lk\, (\mathcal{K}, \,  \mathcal{K}_{\digamma})$ is the self--linking number of the two curves.\\
The freedom in the choice of the framing (actually of the longitude in $\mathcal{N}$)  has to be restricted in order to parametrise in an explicit way the Seifert fibration defined in \eqref{fibra1} and \eqref{fibra2}. The first step consists in setting 
\begin{equation}
	\label{framing3}
   Wr( \mathcal{K})\,=\,Tw  (\mathcal{K}, \, \mathcal{K}_{\digamma})\,\; \Rightarrow\;\,
    Lk\, (\mathcal{K}, \mathcal{K}_{\digamma}) \,=\, 2\, Wr( \mathcal{K}) \,=\,2 \,Tw  (\mathcal{K}, \, \mathcal{K}_{\digamma}),
\end{equation}
a prescription called ``vertical framing" \cite{Rol,Lic}, and commonly adopted  in dealing with link invariants in Chern--Simons topological quantum field theory \cite{Gua}. \\

Looking at the geometric features of both the solid torus $\mathcal{T}$ and the tubular neighbourhood  $\mathcal{N}$, the natural coordinates are obviously the cylindrical ones.  In particular, given that
 $\mathcal{T} \,=\, S^1 \times D^2 $, with  $S^1=\{z_1 \in \mathbb{C}\,| \;|z_1|^2 = 1\}$ $:= \mathbb{C}^* $ and $D^2 = \{z_2 \in \mathbb{C} \,|\; |z_2|^2 \leq 1\}$,  the complex parametrisation reads
\begin{equation}
(\exp (2\pi i \lambda), \, r \, \exp (2\pi i \mu)) 
\,:=\, (z_1, z_2) \, , 
\label{torus1}
\end{equation}
where 
$r \in (0,1],\, \lambda \in [0,1],\, \mu \in [0,1]$ and
$|z_1|^2 + |z_2|^2 = 1+r^2 \subset \mathbb{R}^4$.  Indeed, since the boundary $\partial\mathcal{T}$ of  
$\mathcal{T}$ is the 2-dimensional torus $T^2= S^1 \times S^1$  parametrised by 
$(\exp (2\pi i \lambda),\,\exp (2\pi i \mu))$, the names of the angular variables permit to grasp at once that $(\exp (2\pi i \lambda),1)$ represents a longitude and $(1, \exp (2\pi i \mu))$  a meridian in $T^2$, \textit{cf.} Fig. \ref{Torus}. The pair of oriented curves  
$(\boldsymbol{\lambda}, \boldsymbol{\mu})$  are the generators of the first homotopy group of $T^2$, $\pi_1 (T^2)$, isomorphic to the first homology group with integer coefficients, $H_1 (T^2, \mathbb{Z})$. Any other homotopy class of curves on $T^2$ is expressed in this \textit{Rolfsen basis} \cite{Rol} as
\begin{equation}
	\label{framing5}
		[ \boldsymbol{\gamma}]\,=\,a\, [\boldsymbol{\lambda}]\,+\, b \,[\boldsymbol{\mu}\,]\,;\;\; a,b \in 
		\mathbb{Z}\,.
\end{equation}
Then the self--homeomorphisms (preserving orientation) 
\begin{equation}
	\label{framing6}
		\tau_{\partial}\,:\, T^2 \longrightarrow \,T^2
\end{equation}
are parametrised by matrices of $SL(2, \mathbb{Z})$, the special linear group of $2 \times 2$
matrices with integer coefficients and unit determinant, actually the mapping class group of the torus up to ambient isotopy. The group of homeomorphisms in \eqref{framing6} can be decomposed into
elementary operations, \textit{the twists}, involving  longitudes $(L)$ and  meridians $(M)$, respectively
\begin{align}
	\label{framing7}
\tau_{\partial}^{\pm (L)}\; 
(\exp (2\pi i \lambda),\,\exp (2\pi i \mu)) &= (\exp (2\pi i (\lambda \pm \mu)),\,\exp (2\pi i \mu))\nonumber \\
\tau_{\partial}^{\pm (M)}\; (\exp (2\pi i \lambda),\,\exp (2\pi i \mu))& = (\exp (2\pi i \lambda),\,\exp (2\pi i (\lambda \pm \mu))\,.
\end{align}
The matrix representations are given by $\tau_{\partial}^{+(L)}$ 
$\rightarrow 
(\begin{smallmatrix}1 & 0 \\ 1 & 1
\end{smallmatrix})$ and 
$\tau_{\partial}^{+(M)}$ 
$\rightarrow 
(\begin{smallmatrix}1 & 1 \\ 0 & 1
\end{smallmatrix})$, and the $\tau_{\partial}^{-}$ correspond to the inverse matrices.\\
For what concerns the solid torus $\mathcal{T}$, its  self--homeomorphisms 
are denoted $\tau : \mathcal{T} \rightarrow \mathcal{T}$, and it can be shown \cite{Rol}
that
 $\tau_{\partial} : T^2 \rightarrow T^2$ can be extended  to the solid torus 
if, and only if, these maps takes any meridian to a meridian. This implies that 
 \begin{equation}
	\label{framing8}
		\tau^{\pm}\,:\, (e^{2\pi i \lambda}\,,r \,e^{2\pi i\mu}) \,\, 
\rightarrow \,\,
	(e^{2\pi i\lambda}\,, r \,e^{2\pi i(\lambda \pm \mu )})\,,
\end{equation}
where the + occurs when $\lambda +\mathsf{n} \mu < 1$ and the minus when 
$\lambda -\mathsf{n} \mu > 1$. Accordingly, 
under such twists the Rolfsen basis must transform as
\begin{align}
	\label{framing9}
\boldsymbol{\lambda}'\,=& \,\boldsymbol{\lambda}\,+ \, \mathsf{n} \,\boldsymbol{\mu}\, ; \;\; \;\; \mathsf{n} \in \mathbb{Z}\nonumber\\
\boldsymbol{\mu}'\,=& \, \boldsymbol{\mu}\,.
\end{align}
These particular maps are just meridian twists representing curves that warp around 
 the solid torus modifying its standard longitude. Denoting  $(\tau)^{\mathsf{n}} $  such twists,
 the associated matrices  $\in SL(2,\mathbb{Z})$ have the form
 \begin{equation}
	\label{framing10}
	(\tau)^{\mathsf{n}} \;\; \rightarrow \; \; 
\begin{pmatrix}	
1 & \mathsf{n} \\
0 & 1
\end{pmatrix} \;\, \in \, F (1)\, \subset \, SL(2,\mathbb{Z})\,,
\end{equation}
with $F (1)$ the free group on a single generator, and this complies with the fact that the first homotopy group of the solid torus is $\mathbb{Z}$.\\
Coming back to the framing map in \eqref{framing1}, in order to extend the twists of the solid torus in \eqref{framing7} to the tubular neighbourhood $\mathcal{N}$ it is sufficient 
to conjugate  them with  $\hat{\Phi}$ 
\begin{equation}
	\label{framing11}
	\hat{\tau}^{\pm}\,:=\, \hat{\Phi} \,\circ \,	\tau^{\pm}\,\circ \,\hat{\Phi}^{-1}
\end{equation}
to get the so--called {\bf Dehn twists}, the basic ingredients in the study of the topology of compact 3-manifolds by surgery operations, \textit{cf.} \cite{Rol,Lic} and \cite{Gua}
in the context of quantum field theory.\\

\noindent {\bf Remark.} It is worth  stressing that  we are not dealing with torus knots --namely  simple strands knots lying on the standard 2-torus $T^2$-- fully characterised in the Rolfsen basis
if the integers  $a,b$ in \eqref{framing5} are relatively primes, {\it cf.} Section 4.1.  
Here the tubular neighbourhood $\mathcal{N}$ is globally knotted,
as depicted in Fig.\ref{tubb}, while the standard solid torus $\mathcal{T}$ in Fig. \ref{Torus} can be viewed as a framed counterpart of the unknot. 

\vfill
\newpage

%%%%%%%%%%%%%%%%%%%%%%%%%%%
%%%%%%%%%%%%%%%%%%%%%%%%%%%%
%%%%%%%%%%%%%%%%%%%%%%%%CAMBIATO TITOLO
%%%%% NUOVA TRATTAZIONE OPTICAL KNOTTED FIELDS
%%%%%%%%%%%%%%%%%%%%%%%%%%%%%%%%%%%%%%%
%%%%%%%%%%%%%%%%%%%%%%%%%%%%%%%%%%%%%%%%%%%

\subsection{Preferred framing and parametrisation of knotted\\ 
optical fields}

On the basis of the results  reported in the previous section, we define the {\bf preferred framing} to be assigned to an oriented, framed knot (or, equivalently, to a band $(\mathcal{K}, \mathcal{K}_{\digamma})$) as the combination of the vertical framing defined in \eqref{framing3}  with the selection of the Rolfsen basis in the associated tubular neighbourhood $\mathcal{N}$ as in \eqref{framing9} supplemented by \eqref{framing11}. 
Accordingly, the topological invariant $\mathsf{n}$ (twice the twisting number of the band) 
in \eqref{framing10}, characterises the knotting features of the framed knot and, choosing the orientation of the longitude $\boldsymbol{\lambda}$ in the anticlockwise (positive) direction as well as  the $\mathsf{n}$ meridians
$\boldsymbol{\mu}$'s,  $\mathsf{n}$ turns out to be a natural number $\geq 1$.\\
With these premises, we introduce an explicit parametrisation of the Seifert fibration  (to the best of our knowledge, so far not  considered in literature).
Denoting $\hat{\tau}^{\mathsf{n}}$ the  twist operations in the preferred framing, 
the original Seifert map $f$ in Eq. \eqref{fibra1} is upgraded to
$\hat{f}^{(\mathsf{n})}:=$ $f \circ  \hat{\tau}^{\mathsf{n}}$, namely  
\begin{equation}
\label{preferred1}
S^1 \times D^2  \xrightarrow{\,\,\hat{\tau}^{\mathsf{n}}} \; \mathcal{N}  
\subset S^3 \;
\xrightarrow{\;f} \; S^1 \,.
\end{equation}
Of course the $\hat{\tau}$'s (positive Dehn twists) depend on the homeomorphism $\hat{\Phi}:
S^1 \times D^2 \rightarrow \mathcal{N}$  of Eq. \eqref{framing1} through conjugation  as in Eq. \eqref{framing11}.  
Up to homotopy, the  explicit parametrisations of the maps in \eqref{preferred1} reads
\begin{equation}
\label{preferred2}
(e^{2\pi i \lambda}\,,r \,e^{2\pi i\mu})  \mapsto (e^{2\pi i \lambda}\,,r \,e^{2\pi i(\lambda + \mathsf{n} \mu)})  \mapsto  e^{2\pi i(\lambda + \mathsf{n} \mu)}. 
\end{equation}
\begin{figure}[ht]
	\includegraphics[height=6cm]{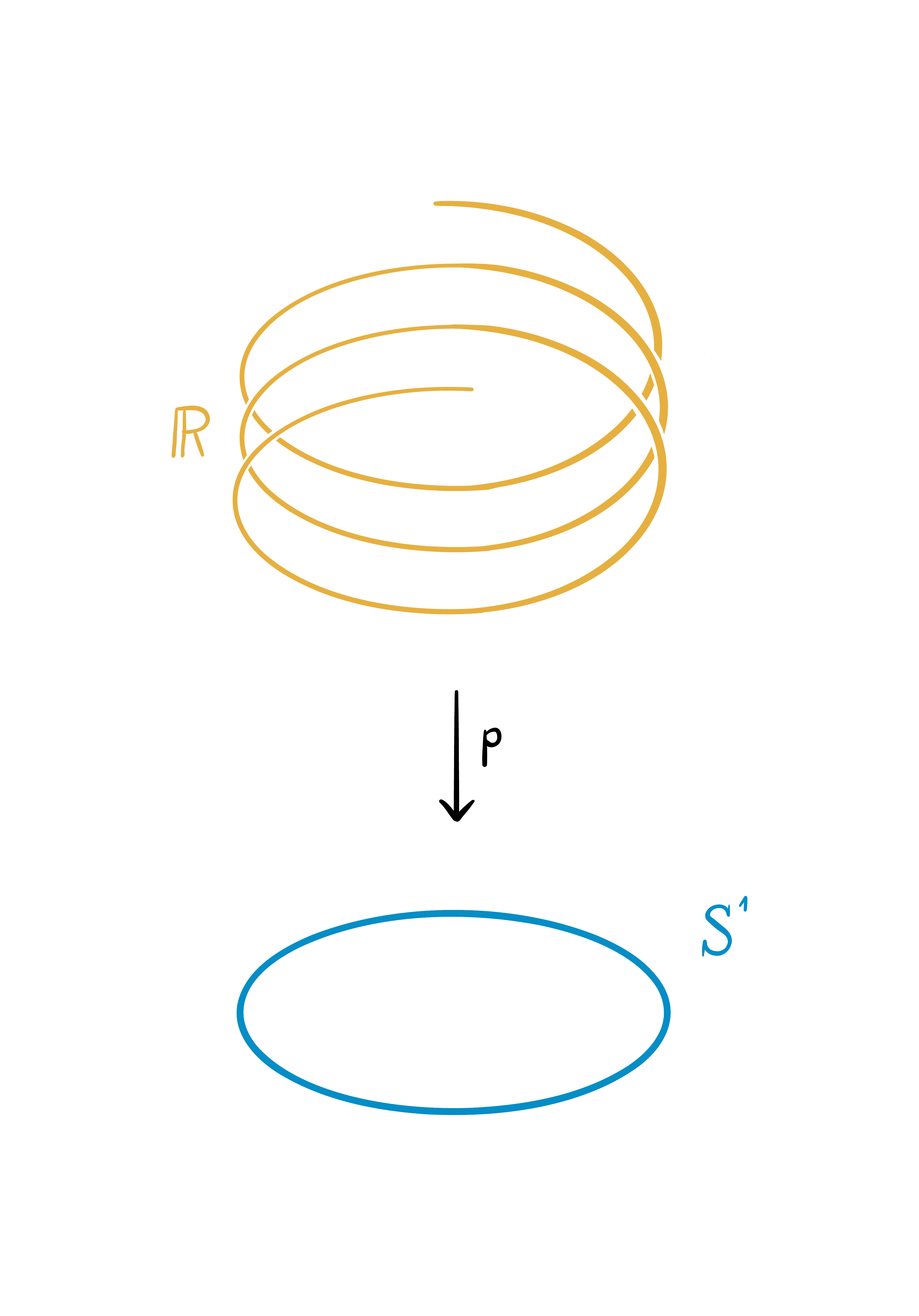}
	\centering
	\caption{The universal covering of the circle.}
\label{Covering}
\end{figure}
The argument in the last exponential, $(\lambda + \mathsf{n} \mu)$ ($\lambda, 
\mu \in [0,1]$), actually varies in the universal covering of the circle $S^1$, the real line $\mathbb{R}$. As shown in Fig.\ref{Covering}  the integer $\mathsf{n}$ can be interpreted as the winding number associated to the projection  $p: \mathbb{R} \rightarrow S^1$.
Recalling that the inverse $f^{-1}$ of the fibration map 
identifies a {\bf Seifert surface}  whose boundary is the knot 
$\mathcal{K}$,
in the preferred framing a particular presentation of the Seifert surface arises, namely
\begin{equation}
\label{preferred3}
\mathbb{R}  \xrightarrow{\,p} \;S^1 \,   \xrightarrow{\;(\,\hat{f}^{(\mathsf{n})}\,)^{-1}} \; \Sigma^{(\mathsf{n})}\;\subset \;\mathcal{N} \,.
\end{equation}
The visualisation of this surface is not easily worked out, and it is worth noting that it is not necessarily the Seifert surface of minimum genus associated to a given knot. This issue will be discussed in Section 4.2.\\

%%%%%%%%%%%%%%%%%%%%%%%%%%%%%%
%%%%%%%%%%%%%DA QUI ALLA FINE DELLA SEZIONE 
%PARTE NUOVA %%%%%%%%%%%%%%%%
%%%%%%%%%%%%%%%%%%%%%%%%%%%%%%%%%%

In order to introduce Ra\~{n}ada--type fields shaped like a knotted solid torus $\mathcal{N}$,
a closer inspection of  the geometric features of this compact oriented 
submanifold  of the smooth differentiable manifold $S^3$ is needed.  
As a preliminary remark, the manifest cylindrical symmetry  permits to equip $\mathcal{N}$ with a  Riemannian structure: 
this metric lives in a neighbourhood of $\mathcal{N}$ where the local cylindrical coordinate introduced in  \eqref{torus1} must be suitably restricted. More precisely, fix a  $\varkappa$ $\in (0, 1]$ and restrict the domain of the radial variable $r$ to $\rho \in (1- \varkappa,1]$. Then the centre of the disk $D^2$ --the intersection with the knot curve-- is necessarily removed, and the
domain $D^2$ becomes an open ``annulus" $\mathcal{A}_{\varkappa}$ as shown in the middle cartoon of Fig.\ref{contract}. Such a cylindrical slice (topologically a solid ``hollow" torus) is parametrised as
\begin{equation}
\label{torus4}
\mathcal{N}_{\varkappa}\,\cong \,S^1 \times \mathcal{A}_{\varkappa}  \,:= \,\{ (\rho, \lambda, \mu)\,\,| \rho \,\in (1- \varkappa,1]\}\,,
\end{equation}
and the (diagonal) metric tensor $\mathsf{g}$ reads
\begin{equation}
\label{torus5}
\mathsf{g}\,=\,g_{\rho \rho}(\rho)\, d\rho \otimes d\rho \,+\,g_{\lambda \lambda}(\rho) \, d\lambda \otimes d\lambda \,+\, g_{\rho \rho} (\rho) \, d\mu \otimes d\mu  \,,
\end{equation}
where the components  are suitable smooth function of the radial variable (more details on the intrinsic and extrinsic geometry of solid tori can be found in \cite{Pul}). \\

%%%%%%%%%%%%%%%%%%%%%%%%
%%%%%%%%%%%%%%%%%%%%%%%%%%%%
With the aim of mimicking the construction of the Hopf--Ra\~{n}ada fields 
({\it cf.} the Introduction) the relevant geometric quantities are indeed 
{\bf area $2$-forms}. Thus it is necessary to made an extension of the codomain of the fibration map defined in Eq. \eqref{preferred1}, 
from $S^1$  to the 2-dimensional set $\mathcal{A}_{\kappa}$. To simplify the notation with respect to $\hat{f}^{(\mathsf{n})}$, the upgraded fibration map becomes  
\begin{equation}
\mathfrak{f}^{\,\mathsf{n}} \,:\,  \mathcal{N} \subset  \,S^3
 \;\rightarrow \; \mathcal{A}_{\varkappa} \,\subset \mathbb{C} \,.
 \label{torus6}
\end{equation}
Coming back to the coordinatisation of the standard 2-disk $D^2 \subset \mathcal{T}$ given in Eq. \eqref{torus1} and setting $z_2 \equiv z = r\, \exp(\,2 \pi i \mu)$, the normalised expression of its area 2-form reads
\begin{equation}
\sigma_{\,D^2}\,=\, 
\frac{1}{2 \pi i} \; \frac{d\bar{z} \wedge dz}{z \bar{z}}\,
\propto \, 
\frac{r dr \wedge d\mu}{r^2}\,.
\label{area1}
\end{equation}
In the preferred framing \eqref{preferred2}  the cylindrical coordinatisation of the annulus $\mathcal{A}_{\varkappa}$ is given by $\rho \in (1- \varkappa,1]$ and  $\lambda + \mathsf{n}\,\mu := \nu$, and the complex counterpart reads
\begin{equation}
\zeta 
\,=\, \rho \, e^{2 \pi i \nu}\,;\;\;  \bar{\zeta} 
\,=\, \rho \, e^{\,-2 \pi i \nu} \,.
\label{area2}
\end{equation} 
The associate normalised area 2-form 
\begin{equation}
\sigma_{\,\mathcal{A}_{\varkappa}}\,=\, 
\frac{1}{2 \pi i} \; \frac{d\bar{\zeta} \wedge d\zeta}{\zeta \bar{\zeta}}\,
\propto \, 
\frac{\rho d\rho \wedge (d\lambda + \mathsf{n} d\mu)}{\rho^2}\,
\label{area3}
\end{equation}
can be pulled back through $(\mathfrak{f}^{\,\mathsf{n}}\,)^*$  $:\,\Lambda^2 (\mathcal{A}_{\varkappa}\,)\,\rightarrow\,$ $\,\Lambda^2 (\mathcal{N}\,)$, the spaces of differential 2-forms on the corresponding manifolds. 
This operation is not affected by obstructions because, on the one hand,  $\mathcal{N}$ and $S^1$ (a deformation retract of  $\mathcal{A}_{\varkappa}\, , \forall \varkappa \,$) have the same homology groups, $H_1 = \mathbb{Z}\,, H_2 = 0$. On the other hand, according to the prescription encoded in the choice of the Rolfsen basis in Eqs. \eqref{framing8} and \eqref{framing9}, the boundary
$\partial \mathcal{A}_{\varkappa}$ must match with the boundary of a meridian disk
$\subset \mathcal{N}$ for each value of the longitude angle $\lambda$. Then
\begin{equation}
(\mathfrak{f}^{\,\mathsf{n}})^* \,(\sigma_{\,\mathcal{A}_{\varkappa}})\,:=\,
\sigma_{D^{(\mathsf{n})}}\,=\, \lim_{\rho \rightarrow r}
\,\sigma_{\,\mathcal{A}_{\varkappa}}\,=\, 
\frac{1}{2 \pi i} \; \frac{d\bar{\boldsymbol{\zeta}} 
\wedge d\boldsymbol{\zeta}}{\boldsymbol{\zeta} \bar{\boldsymbol{\zeta}}}\,
\propto \, 
\frac{r dr \wedge (d\lambda + \mathsf{n} d\mu)}{r^2}\,
\label{area4}
\end{equation}
represents the area form of an $\mathsf{n}$--twisted meridian disk in $\mathcal{N}$ or, equivalently, of a slice of the Seifert surface 
$\Sigma^{(\,\mathsf{n})}$ in \eqref{preferred3} whose height is $d\lambda$.
Following the remarks on the Ra\~{n}ada construction in the Introduction, 
the  {\bf Seifert--Ra\~{n}ada} fields, denoted $\Gamma (\mathbf{r}\,,t)$ 
and $\Xi (\mathbf{r}\,,t)$, have to share the same expression with the Faraday 2-forms $\mathsf{F}$ and $^*(\mathsf{F})$, and this happens  once   we replace in Eq. \eqref{area4}
$\boldsymbol{\zeta}$ with $\Gamma$ and $\Xi$ (and their complex conjugates), respectively.\\
We are not going to discuss in this paper the physical properties of these complex fields (this issue would deserve further investigations).
Instead, we want to focus on a few more geometric remarks. In the first place, it is clear that the singularity in expression \eqref{area4} occurs at $r=0$, 
namely where the disks intersect the core knot $\mathcal{K}$. This is of course different from what happens for the Hopf map (and Hopf--Ra\~{n}ada  fields) where projective coordinates are singular at the South pole of the 3-sphere $S^3$. Moreover, it is crucial to check the compactification 
conditions on domain and codomain of the improved Seifert map 
\eqref{preferred1}. As said before, the knotted solid torus $\mathcal{N}$ is a compact submanifolds of $S^3$ and the conditions
\begin{equation}
\lim_{| \mathbf r | \rightarrow \infty}\:
\Gamma (\mathbf{r}\,,t) \,=0   \;\; \text{and}\;\;
\lim_{| \mathbf r | \rightarrow \infty}
\Xi (\mathbf{r}\,,t) \,=0 
\label{area5} 
\end{equation}
can be fulfilled by looking at $r$ (the radial cylindrical coordinate in $\mathcal{N}$)
as indistinguishable from the radius of $S^3$ in the limit. The codomain of 
$\mathfrak{f}^{\,\mathsf{n}}$, $\mathcal{A}_{\varkappa}$, retracts naturally 
on its boundary $S^1 \cong \mathbb{C}^* $ (unit complex numbers) and thus the inverse map
does not depend either on the direction approaching to infinity in $\mathbb{C}$.

%\vfill
%\newpage

\section{From braids to braided open books}
\begin{figure}[ht]
	\includegraphics[width=6cm]{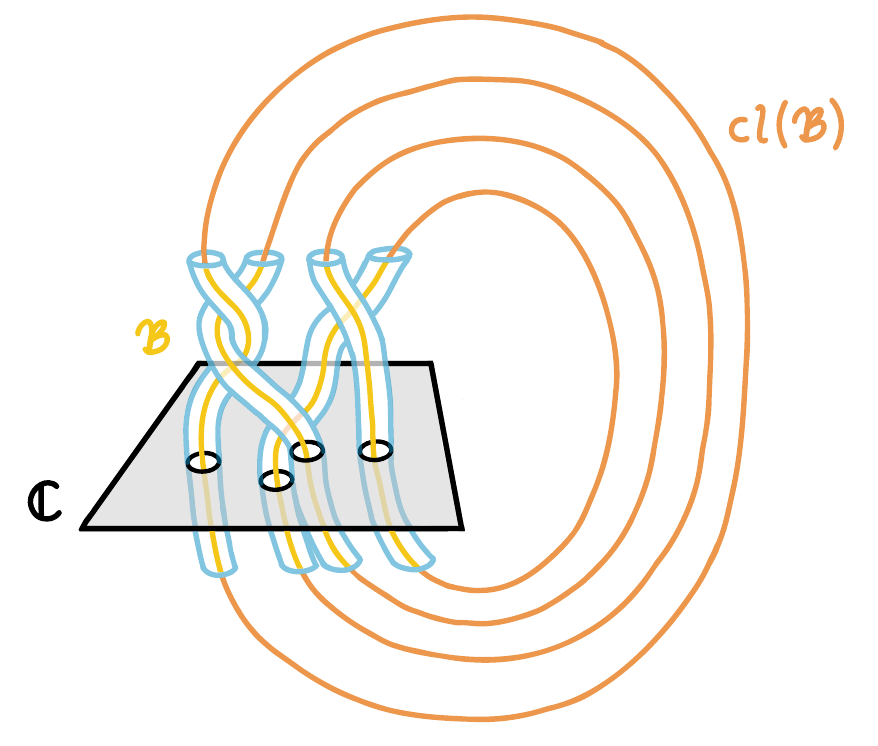}
	\centering
	\caption{The orange curves represent the closure $cl(\mathcal{B})$ of a collection of braided strands, $\mathcal{B}$, and their intersection with a 2-disk 
is shown. The knottiness properties of the tubular neighbourhood can be best addressed by resorting to braided open books.}
\label{tub}
\end{figure}

The close relation between braids and knots or links (here considered as knotted and linked simple closed curves) was discovered by Alexander in \cite{Ale1}. Given any knot
$\mathcal{K}$  in $\mathbb{R}^3$ or $S^3$ there exists a (not unique) braid $\mathcal{B}$ 
such that $\mathcal{K}$ is its  ``closure", as depicted in Fig.\ref{tub}. This transformation is efficient, that is, there exists an algorithm that may run in polynomial time to achieve the result. 

\noindent Beside the crucial role played by the braid group in mathematical knot theory 
it has been recognised also in  experimental works that, in order to study knotted structures, it is useful to consider their ``presentations" as  braids --not to be confused with ``representations"-- which refer to  linear maps of the braid group in other  algebraic structures.

\subsection{The braid group B$_n$}
\begin{figure}[ht]
	\includegraphics[width=8cm]{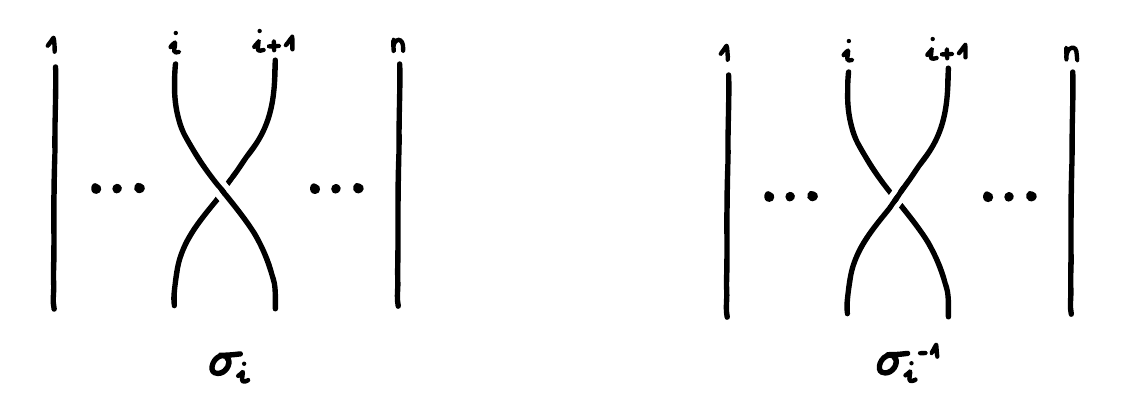}
	\centering
\caption{The $\sigma_i$ generator acting on the $i^{th}$ strand, creating an over-crossing of $i$. The corresponding under-crossing is  given by the $\sigma^{-1}_i$ generator. The orientation is from top to bottom.}
\label{generators}
\end{figure}

\begin{figure}[ht]
	\includegraphics[height=6cm]{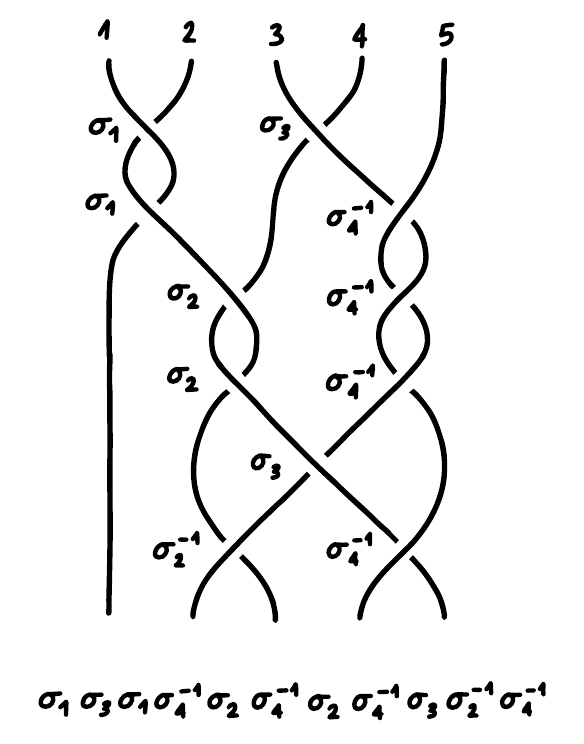}
	\centering
	\caption{An example of a braid on five strands and its corresponding braid word. The generators are listed from top to bottom.}
	\label{br1}
\end{figure}

Following in this section the review paper \cite{BiBr}
a {\bf geometric  braid} on \textit{n} strands is characterised by drawing \textit{n} strings hanging from the top at \textit{n} distinguished locations in a cylinder $D^2 \times [0,1]$ and building up the whole braid by iteratively applying over-- and under--crossings, from bottom to top. The elementary crossings,  denoted  $\sigma_i$ (for $i=1,2,...,n-1$), are operations acting on  two contiguous  strands of the braid, as depicted in 
Fig.\ref{generators}.  Each braid can be encoded into a \textit{braid word} by composing the elementary operations 
$\left\lbrace \sigma_1, \sigma_2, \ldots, \sigma_{n-1}\right\rbrace$, 
which are --adding the identity $e$-- the \textit{n generators} of the Artin braid group on $n$ strands, B$_n$. A \textit{word} of length \emph{k} is a concatenation of \emph{k} $\sigma_i$ and $\sigma^{-1}_i$. The composition of the generators is meant such that the crossings are performed from left to right in the braid word, and from top to bottom in the braid diagrams, see an example in Fig.\ref{br1}.

\begin{figure}[ht]
	\includegraphics[width=7cm]{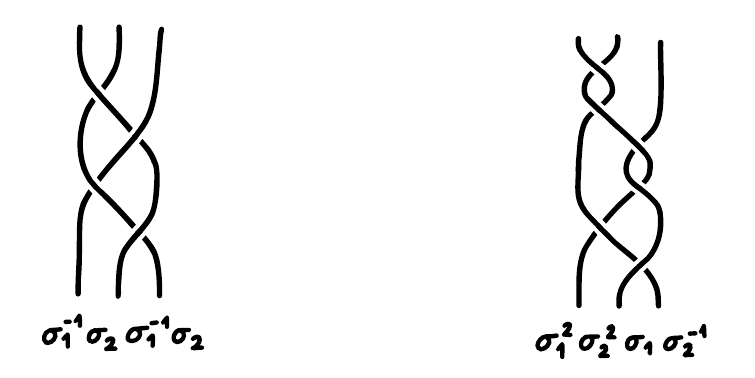}
	\centering
	\caption{A homogeneous braid (left) and a non--homogeneous braid (right)}
	\label{hom}
\end{figure}

\noindent The generators of B$_n$ must  satisfy the following \textit{relations}\\
$\bullet$ commutation under exchanges of non--consecutive generators
		\begin{equation}
		\label{braid1}
			\sigma_i \sigma_j=\sigma_j\sigma_i \; \text{when} \;\lvert i-j \lvert \geq2 \; 
\text{for} \; i,j \in \left\lbrace 1, \ldots, n-1\right\rbrace 
		\end{equation}
$\bullet$ a three--term relation (of Yang--Baxter type)  under swapping of consecutive generators 
		\begin{equation}
			\sigma_i \sigma_{i+1}\sigma_i=\sigma_{i+1}\sigma_i\sigma_{i+1}\; \text{for} \; i\in \left\lbrace 1, \ldots, n-2\right\rbrace 
			\label{braid2}
		\end{equation}
A braid word corresponding to $\mathcal{B}$ is \textbf{homogeneous} if\\
	$\bullet$  each generator $\sigma_i$ appears at least once;\\
$\bullet$  each exponent of $\sigma_i$ has the same sign and no other exponent is allowed,
 see Fig.\ref{hom}.
As already noted, the  presentation of a knot $\mathcal{K}$ as the closure of a braid, see Fig.\ref{tub},  is by no means unique, and moreover the problem of finding out a braid with the minimum number of strands for a given knot is in the class of NP problems. Such a minimum number is called the \textbf{braid index} of the knot (link). In order to understand if two braids do represent the same knot, one has to check whether it is possible to go from one to the other by performing a (finite) sequence of the so--called \textit{Markov moves}. These operations on braids are basically the algebraic counterpart of the \textit{Reidemeister moves} on knot diagrams that are used to prove if two knots are actually ambient isotopic. A few more details on this subject are collected in Appendix B.

%%%%%%%%%%%%%%%%%%%%%%%%%%%%%
%%%%%%%%%%%%%%%%%%%%%%%%%%%%%%%%%%%%%%%%%%%%%%%%%
%\vfill
%\newpage

\subsection{Open book decomposition and classes of fibrable knots}
Generally speaking, open book structures provide decompositions of some given $n$-dimensional compact and oriented manifold into compact submanifolds of lesser dimension. As is well--known \cite{Rol}
oriented fibred knots (links) in the 3-sphere $S^3$ are particular instances of \textbf{open books}, and their definition coincides with that of the Seifert fibration map $f$ introduced in Section 2.1. It is not necessary to modify  the notation used there; just changes of names are in order: the knot $\mathcal{K}$
becomes the ``binding", and for every $\theta \in [0,1]$, the fibres $f^{-1}(\theta)$ are the ``pages" of the book, whose  collection represents a Seifert surface $\Sigma$  with boundary the binding. Moreover, in order to point out the open book content of the Seifert fibration map we will use the notation $(\mathcal{K}, f)$.\\
Coming back to the question ``can all knot types be fibrable?", in the mathematical literature 
\cite{Har,Sta} it has been reported that fibrable knots are all the torus knots, the figure--eight knot, the square and the granny knots, and the Turk's head knots
\cite{Knots}. Moreover, all knots (links) that can be presented as  homogeneous braids are fibrable. 
%%%%%%%%%%%%%%%%%%%%%%%%%%%%%%%%%%%%%%%%%%%%
%%%%%%%%%%%%%%%%%%%%%%%%%%%%%%%%%%%%%%%%%%%%%%%%
%%%%%%%%%%%%%%%%%%%%%%%%%%%%%%%%%%%%%%%%%%%%%%

\subsection{Braided open books}
Having established a direct relationship between open books and fibred (framed) knots, we now introduce the concept of \textit{braided} open books, thus shifting the focus to fibred braids as the main objects of interest. As a consequence of Alexander's theorem, given an open book decomposition of $S^3$, it is always possible to find a (non-unique) braid $\mathcal{B}$ whose closure  $cl\,(\mathcal{B})$ is the binding of the corresponding open book $(\mathcal{K}, f)$.
However, such a  na\"ive construction  turns out to be restrictive for our purposes and not quite effective.  Different ways of braiding open books are reviewed  in details in the Thesis 
\cite{Tesi}. We present here just a summary of those results that set explicit constraints on classes of braids, noting that in \cite{Bode} it has been shown that the different notions of braidability are equivalent.

\noindent {\bf Open books from generalised exchangeable braids}\\
This braiding prescription, first proposed in \cite{Mor}, consists in 
considering an open book decomposition, namely $(cl(\mathcal{B}), f)$, where $cl(\mathcal{B})=\mathcal{K}$ is the closure of a braid $\mathcal{B}$.
Whenever $\mathcal{K}$ is positively transverse to the pages of ($\mathcal{O}$, $f$) --the unbook--, and $\mathcal{O}$ --the unknot-- is positively transverse to the pages of ($cl(\mathcal{B})$, $f$), it is possible to obtain a braided open book. \\
Recall that the braid index of a given knot is the \emph{minimum} number of strands required to express such knot as a closed braid.
According to \cite{Rud1,Rud2,Bode}:
\begin{quote}
{\bf 1)} whenever the binding $\mathcal{K}$ has braid index less than or equal to 3, then $\mathcal{K}$ can be expressed as the closure of a generalised exchangeable braid.
This implies that, given an open book ($\mathcal{K},f$), whose binding has braid index of at most 3, then the open book can always be braided.
\end{quote}

\begin{figure}[ht]
	\includegraphics[width=8cm]{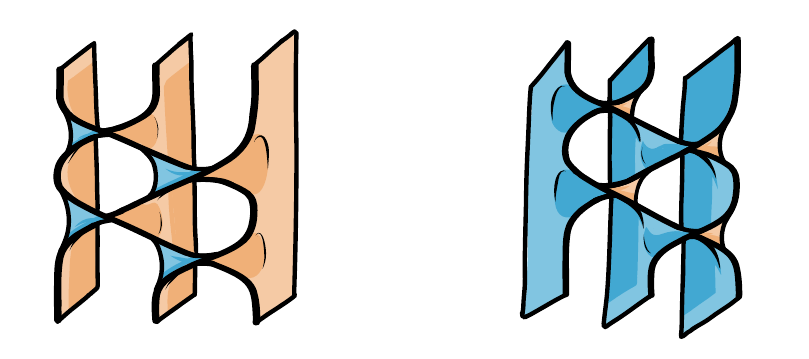}
	\centering
	\caption{Two views of the Seifert surface of a homogeneous braid: a top--right corner view (left) and left--bottom corner (right). The two sides of the surface have been coloured differently in order to better distinguish the twists.}
	\label{brd}
\end{figure}

\noindent {\bf Open books from homogeneous braids}
This issue would call into play the notion of P-fibred braids, not given here explicitly. Actually, it has been shown that several families of fibred knots can be obtained by considering the closure of P-fibred braids 
(see \cite{Bode} and references therein). On the other hand, it is not known whether every fibred knot is the closure of a P-fibred braid. Nevertheless, 
\begin{quote}
{\bf 2)} there exist  particular classes of braids (on $\geq 2$ strands) that are always P-fibred, the homogeneous braids defined at the end of the previous section, \textit{cf.} Fig.\ref{hom}. 
\end{quote}
\noindent Denoting $\mathcal{B}^{hom}$ a homogeneous braid, it turns out that an associated  Seifert surface $\Sigma$ $(cl(\mathcal{B}^{hom}))$ can be obtained from the union of \emph{n} disks, one for each strand, where the $i^{th}$ and the $(i+1)^{th}$ disks are joined by a number of right- (left-) handed half--twisted strips, one for each occurrence $\sigma_i$ ($\sigma^{-1}_i$, respectively) in $\mathcal{B}^{hom}$. The pictorial representation of this kind of Seifert surface, before the closure of the binding,  is  visualised in Fig.\ref{brd}.

%\vfill
%\newpage

%%%%%%%%%%%%%%%%%%%%%%%%%SECTION 4
%%%%%%%%%%%%%%%%%%%%%%%%%%%%%%%%%
%%%%%%%%%%%%%%%%%%%%%%%%%%%%%%%%%%%%

\section{Topological invariants of Seifert surfaces and \\ knotted framed configurations}
In this section we summarise and upgrade what has been found in the previous sections about the topological characterisation of the Seifert fibration and associated framed knots and braided open books. Given the pivotal role played by (different presentations of) Seifert surfaces anywhere in the previous sections, we give in Section 4.2 a short overview on the basic topological features of Seifert surfaces, and  we identify in the Alexander--Conway polynomial a useful invariant for  characterising  (optical) knotted framed configurations (Section 4.3).
Throughout the text  emphasis is put on algorithmic and computational questions.

\subsection{Numerical invariants and constraints}
As reported in Section 2.3, the self--linking number, a topological invariant  of a framed knot presented as a band $(\mathcal{K}, \mathcal{K}_{\digamma})$,  is given by twice the number of positive twists, $2\, Tw (\mathcal{K}, \mathcal{K}_{\digamma})$ $\equiv 2 \mathsf{n}$, once the preferred framing is adopted (\textit{i.e.} the vertical framing in the Rolfsen basis). This integer provides what we have called a ``parametrisation" of the associated Seifert surface $\Sigma^{(\mathsf{n})}$, but of course does not provide much  information on the overall topology of the surface.

Coming to the algebraic setting of Section 3, it is clear that the braid index cannot represent an invariant of the associated knot (link), still it is equal to the minimum number of Seifert circles \cite{Yam} that, in turn, arise as basic ingredients of the Seifert algorithm (explained below). Moreover, as already noted, there are no efficient algorithms able to compute the braid index, a problem strictly related to the conjugacy problem  in the braid group B$_n$. If $\mathcal{W}$ is the word  associated to some braid  $\mathcal{B}$, a conjugate braid is given by the word $\alpha \,\mathcal{B}\,\alpha^{-1}$, with $\alpha$ one of the generators $\{\sigma_1, \sigma_2, \dots, \sigma_{n-1}\}$. Then the best  known algorithm for deciding whether the two braids are equivalent is exponential in both $n$ and  the length $|k|$ of the braid word $\mathcal{W}$. Such types of NP--problems are quite common in combinatorial group theory, and we refer to Section 5 of \cite{BiBr} for details and references. \\

\noindent {\bf Remark.} The constraint on the braid index given in item 1) of Section 3.3
(\textit{if the binding  $\mathcal{K}$ of an open book has braid index $\leq 3$ then the open book can always be baided}) actually complies with the experimental settings implemented so far. The light beams are manipulated as to draw the trefoil, cinquefoil and figure--eight knots. Knots and links which are closed 3-braids share a number of interesting features; a classification theorem can be found in \cite{BiMe}. $\blacksquare$\\

\begin{figure}[H]
	\includegraphics[width=8cm]{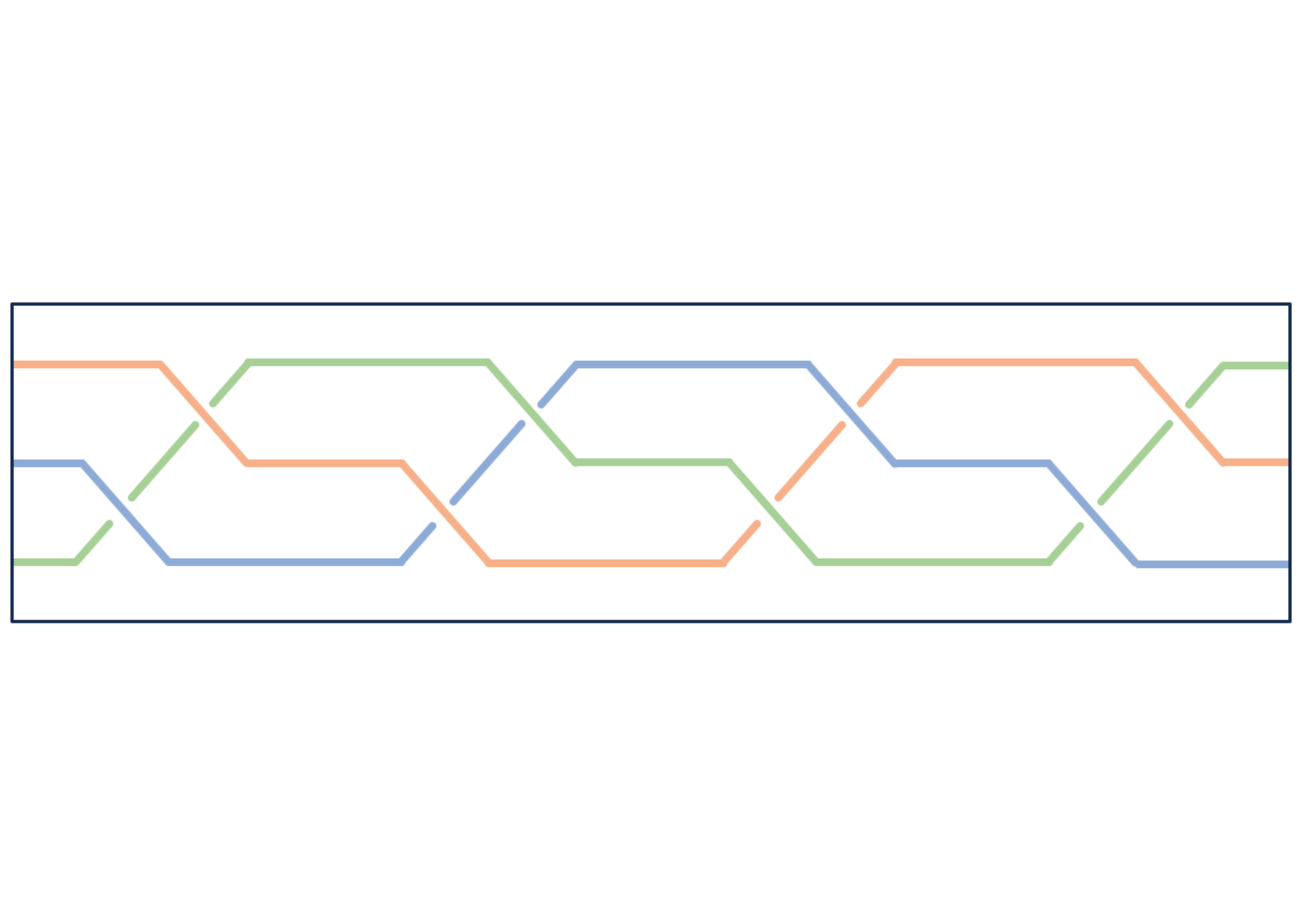}
	\centering
	\caption{The positive homogeneous braid whose closure is the torus knot
$(3,4)$}
	\label{Hombraids}
\end{figure}

\begin{figure}[H]
	\includegraphics[width=8cm]{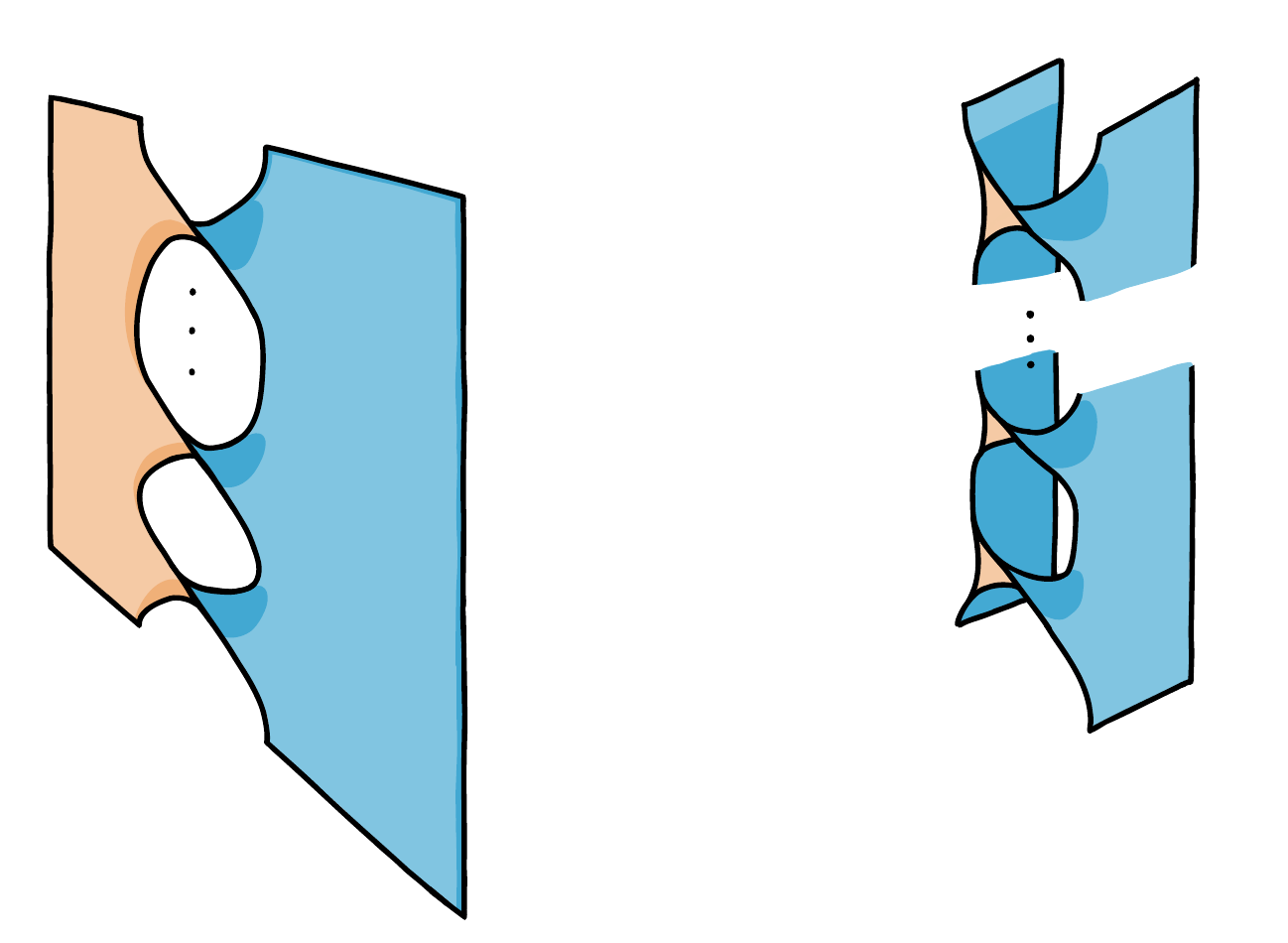} 
	\centering
	\caption{Two pictorial representations of Seifert surfaces associated to homogeneous braids (orange or blue, depending on the orientations). The one on the left results useful only when the braid index is 2; the one on the right is obtained by bending the sides of the surface into the third dimension, allowing the visualisation of structures with braid index greater than 2 upon composition of several blocks.}
	\label{surf}
\end{figure}

\noindent Numerical simulations and experiments reported in \cite{KeBiPe,LaSuMo} deal with optical \textit{torus knots} of type $(p,q)$,
--$p,q$ relatively prime--, where the beams are supposed to warp around the intersection of the Clifford tori of the Hopf fibration (Appendix A).  Recalling  what has been reported at the beginning of Section 3.2, these knots can indeed be fibred, a result found in the context of the topological definition of Seifert fibration, and thus might  be looked at as knotted solid tori, see also the remark at the end of Section 2.2.\\
To be defined, let us remind the  definition of this important class of  knots. If $T^2$ is a standard (unknotted) 2-torus embedded in $S^3$, a $(p,q)$ knot admits a standard presentation in terms of the Rolfsen basis of Eq.\eqref{framing5} as
$p\, [\boldsymbol{\lambda}]\,+\, q \,[\boldsymbol{\mu}\,]$. The appropriate braid word is given by \cite{Lic}
\begin{equation}
			\left(\,\sigma_1 \sigma_2 \cdots \sigma_{p-1}\,\right)^q \,,
			\label{torknot}
\end{equation}
with braid index $p$, and by convention each $\sigma_j$ is a right (positive) twist (note that the trefoil and the cinquefoil knots are toric, of type $(2,3)$ and $(2,5)$, respectively).
In this expression we recognise  explicit examples of \textbf{positive homogeneous braids}, a subclass of homogeneous braids where the power of each generator is $+1$, see an example in Fig.\ref{Hombraids}. According to Section 3.3 these braids are always associated to braided open books.  
The last  remark  should have highlighted the role of these structures  on both the mathematical side and  as useful tools for the practical manipulation of knotted optical configurations. The latter are of course build up by means of  controlled sequences of under- and over-crossings, and it should be feasible to ensure the homogeneity condition at each step, even in cases
where more than two beams could be effectively handled.  \\
The rendering of  the Seifert surface for a homogeneous braid  in Fig.\ref{brd} of Section 3.3, can be improved as shown in Fig.\ref{surf}.
Positive homogeneous braid (indeed, those depicted in the two figures) have Seifert surfaces parametrised by the number of generators in the braid word, namely by the number of positive twists $k$, giving some $\Sigma^{(\,k)}$. As discussed at the beginning of this section, the Seifert surface associated to the band $(\mathcal{K}, \mathcal{K}_{\digamma})$ in the preferred framing is still parametrised by the number of (positive) twists. Nonetheless, the implementation of the preferred framing in the braid context is not feasible: for instance, torus knots would be presented  with just one longitude, so that $p=1$ in Eq. \eqref{torknot} and the braid group would be trivial. According to this remark, the parametrisation with $k$ of the various ``slices" of the Seifert surface in general braided open books does not give an invariant. One should rather use Eq.\eqref{framing2} where the self--linking number is the sum of the weighed crossing number of a planar diagram of the closed braid and of the number of twists $k$.

\subsection{Basics on Seifert surfaces}

\begin{figure}[ht]
	\includegraphics[width=8cm]{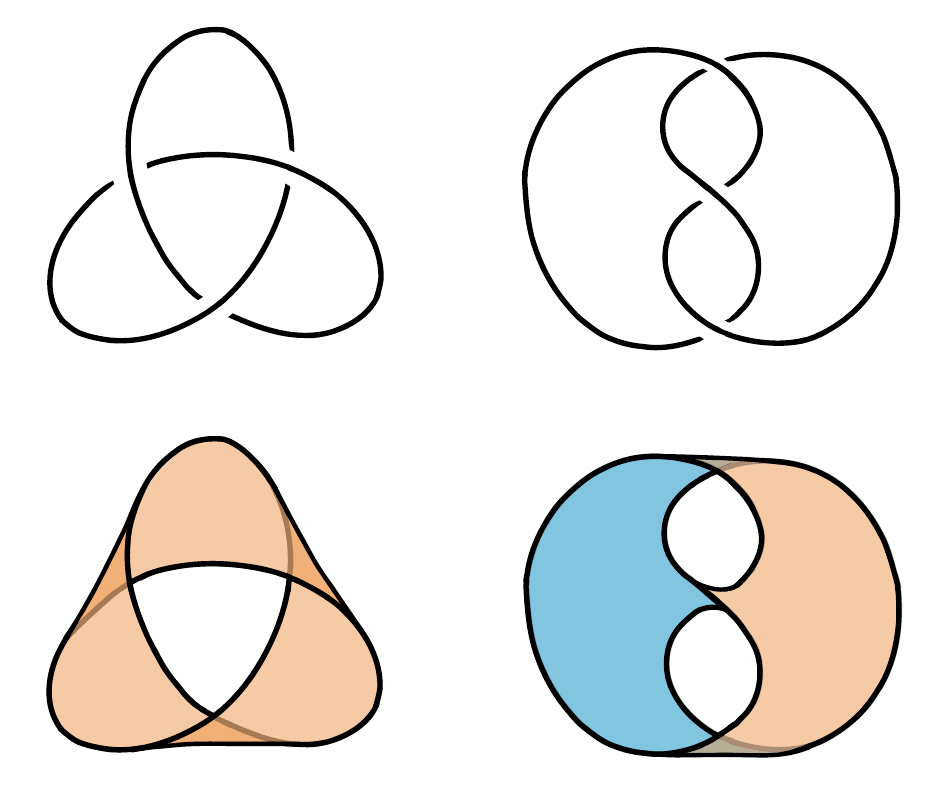}
	\centering
	\caption{Two planar diagrams of the  trefoil knot (top) and their Seifert surfaces (bottom).}
	\label{seifert}
\end{figure}

Recall that the topological type of an oriented, closed surface $\mathcal{S}$ is uniquely determined by its \emph{Euler characteristics} $\chi$, combinatorially computable from a triangulation of $\mathcal{S}$ as $\chi = V-E+F$, where $V$ is the number of vertices, $E$ is the number of edges, and $F$ is the number of triangular faces of $\mathcal{S}$. Alternatively, by resorting to  singular homology theory, $\chi$ can be expressed as the alternating sum $b_0-b_1+b_2$ of the Betti numbers, the ranks of the $m^{th}$ homology groups $H_m$, $m=0, 1, 2$. The invariant $\chi$ is related to the \emph{genus} --that is, the number of holes $g$-- of $\mathcal{S}$ by the  formula $\chi (\mathcal{S})= 2-2g$. In case of compact, oriented surfaces with non--empty boundary $\partial \mathcal{S} \neq \emptyset$, the formula reads
$\chi(\mathcal{S}, \partial\mathcal{S})= 2-2g- \beta$,
where $\beta$ is the number of boundary components. Here $(2-2g)$ is the Euler characteristics of the closed surface obtained from $(\mathcal{S}, \partial \mathcal{S})$ by capping off its boundary components with a 2-disk.\\
In this section,  let $\mathcal{S}(\mathcal{K})$ denote the Seifert surface associated with a fibred oriented knot $\mathcal{K}\subset S^3$, or  with an open book decomposition $(\mathcal{K}, \, f)$ of $S^3$. Then, as explained at length in Sec. 2.1, $\mathcal{S}(\mathcal{K})$ is generated through the inverse of the fibration map $f: S^3 \backslash \mathcal{K} \rightarrow S^1$  and possesses two key properties:\\
$\bullet$
$\mathcal{S}(\mathcal{K})$ is orientable and we agree on a choice of orientation which is compatible with the orientation of $\mathcal{K}$;\\
$\bullet$ $\mathcal{S}(\mathcal{K})$ has a single boundary component given by $\mathcal{K}$ itself.\\
Consequently, its  Euler number is
\begin{equation}
\chi \big(\mathcal{S}(\mathcal{K})\big)=2-2g-1=1-2g\,.
\label{chi}
\end{equation}
In topological knot theory, Seifert surfaces are usually constructed from knot diagrams, which we recall are projection of knots onto a plane. This method, known as {\bf Seifert algorithm}, produces a series of Seifert circles once each crossing is reduced to a pair of arcs. At the end, the Seifert circles are joined together by half--twisted bands and generate the Seifert surface. Without going into details, we just report a few interesting  results, \textit{cf.} \cite{Rol,Lic,Yam}.
Given a Seifert surface $\mathcal{S}$,  there exists a unique knot representing its  boundary.
The converse is not true in general: given a knot $\mathcal{K}$, there can be many Seifert surfaces, depending on the diagram used as the starting configuration in the Seifert algorithm, see Fig.\ref{seifert}.
The \textbf{genus of a knot} is defined as the minimum genus among its Seifert surfaces.
It remains an open problem finding  a way to determine the genus of any given knot. This can be done for some specific types of knots, namely alternating projections of alternating knots or links, and it is known that the genus of a $(p,q)$ torus knot is $g=(p-1)(q-1)/2$.\\
On the algebraic side
recall once more that the minimum number of Seifert circles in any diagram of a knot or link is equal to its braid index. In conclusion, there is no efficient algorithm to compute the genus (or equivalently, the braid index) of a given knot. Moreover, it is known that the genus alone cannot be used to distinguish different knots in general. For example, there are two (fibred) knots of genus one, the trefoil and the (non toric) figure--eight knot,  which are clearly not equivalent \cite{Knots}.

%\vfill
%\newpage

\subsection{The Alexander polynomial invariant}
This quantity, discovered by Alexander in 1923 \cite{Ale2}, is historically the first ambient isotopy polynomial invariant of knots and links (here, collections of simple, closed, oriented curves in $\mathbb{R}^3$
or $S^3$). The discovery in 1985 of the famous Jones polynomial \cite{Jon85} --able to distinguish a knot from its mirror image-- paved the way for the search of other (1- or 2-variable) polynomials of knots, as well as 
invariants of compact 3-manifolds presented as complements of knots in the 3-sphere.
Surprisingly, a few years later Witten proved \cite{Wit} that the Jones polynomial is related to expectation values of
of suitable observables in 3-dimensional topological quantum field theory of Chern--Simons type (the Wilson loop operators).\\
Referring to classic textbooks and review papers \cite{Rol,Lic,Kau} and references therein, we first introduce the Alexander invariant in the homological setting for (connected) Seifert surfaces. This definition encapsulates more information about the topology of an $\mathcal{S} \subset S^3$ with $\partial \mathcal{S} = \mathcal{K}$ (we keep on using the notation of Section 4.2). If $g$ is the (non necessarily minimum) genus, the first homology group of $\mathcal{S}$ with integer coefficients, 
$H_1 (\mathcal{S},  \mathbb{Z})$, is generated by $2g$ closed oriented curves $\{\ell_i\}$, $i=1,2,\dots 2g$.
Since the embedding $\mathcal{S} \subset S^3$ can be extended as to give rise to a ``collar'' 
$\mathcal{S} \times [0,1]$, the curves can be push out in the positive normal direction to 
$\mathcal{S} \times 1$. These images, $\{\ell_j^+\}$, $j=1,2,\dots 2g$, are linked to the $\{\ell_i\}$ 
(of course in each set the curves might be linked among themselves). The resulting configuration of 
$4g$ curves is characterised by the so--called {\bf Seifert matrix} $\mathsf{M}$
\begin{equation}
			\mathsf{M} (\mathcal{S})\,=\, \big( m_{ij} \big)\,:=\, Lk \big(\ell_i, \ell_j^+ \big)\,,
			\label{smatrix}
\end{equation}
with $Lk$ the  linking numbers $\in \{0,+1,-1\}$. Taken a formal variable (an indeterminate) $t$, the Alexander polynomial of $\partial \mathcal{S} = \mathcal{K}$ is a Laurent polynomial in the ring
$\mathbb{Z} \,[t, t^{-1}]$ defined as
\begin{equation}
			\Delta_{\mathcal{K}}\, (t)\,:=\, \det \big(\mathsf{M} - t \, \mathsf{M}^T \big)\,,
			\label{alex}
\end{equation}
where $\mathsf{M}^T$ is the transposed of $\mathsf{M}$ and by convention the polynomial for the unknot $\bigcirc$ is $\Delta_{\bigcirc} = 1$, 
More detail on this construction can be be found in the
the references quoted at the beginning of this section. The key-point is that $\Delta_{\mathcal{K}}\, (t)$ is invariant under operations on the matrix that mimic the Reidemeister moves I,II,III, and thus is an ambient isotopy invariant (see Appendix B).\\

\begin{figure}[H]
	\includegraphics[width=7cm]{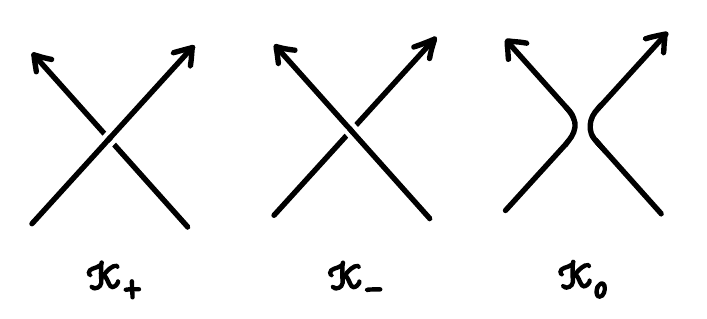}
	\centering
	\caption{Elementary configurations in the skein relation.}
	\label{incroci}
\end{figure}

In the late 60s Conway \cite{Con} found out a recursive procedure to calculate the Alexander polynomial, starting from oriented knot diagrams, on applying a \textit{skein relation}, a three-term expression containing the elementary, local configurations depicted in Fig.\ref{incroci}.
Denoting $w$ the formal variable, the Conway polynomial in the ring $\mathbb{Z}\,[w]$ can be recursively computed from 
\begin{equation}
\nabla_{\mathcal{K}_+} - \nabla_{\mathcal{K}_-} \, =\ w  \nabla_{\mathcal{K}_{0} }
\label{conw}
\end{equation}
with the normalisation $\nabla_{\bigcirc}=1$. The Alexander polynomial can be recovered by the substitution
$ w = (\sqrt{t} -1)/  \sqrt{t}$, and sometimes one refers to the Alexander--Conway invariant to include both parametrisations.\\
A few more properties of this invariant --highlighting its twofold role as a combinatorial object
computed from knot diagrams as well as a quantity able to provide (some) information on the topology of Seifert surfaces-- are collected in the second part of Appendix B. (There we will report also on other knot invariants whose possible role in the physical context will be discussed in the Conclusions.) \\
Note  that in \cite{KeBiPe,ArTiTr} the Seifert surfaces pertinent to specific torus knots solutions have been observed and reconstructed by simulations. It should be possible to compute the corresponding Alexander polynomials and compare the results with known standard expressions of the invariant --for instance, the expression for the trefoil is given in item {\bf i)} of Appendix B.

%%%%%%%%PARTE SUGLI INVARIANTI DEI NODI: SPOSTATA IN APPENDICE B
%%%%%%%%%%%%%%%%%%%%%%%%%%%%%%%%%%%%%%%%%%%%%%%%%%%%%%%%%
%%%%%%%%%%%%%%%%%%%%%%%%%%%%%%%%%%%%%%%%%%%%%%%%%%%%%%%%%%

%\vfill\newpage

%%%%%%%%%%%%%%%%%%%%%%%%%%%%%
%%%%%%%%%%%%%%%%%%%%%%%%%%%%

%%%%%%%%%%%%%%%%%%%%SECTION 5

%%%%%%%%%%%%%%%%%%%%%%%%%%%%%
%%%%%%%%%%%%%%%%%%%%%%%%%%

\section{Conclusions and outlook}
%%%%%%%%%%%%%%%%%%%%%%%%%%%%%%%%%%%%%%%%
As stressed in the Introduction, the aim of this paper has been to highlight topological and geometric structures associated to knotted optical fields going beyond those considered 
so far in the
literature (the Ra\~{n}ada Hopfion solution and its generalisations to solutions representing torus knots, \textit{cf.} the remark at the end of Section 2.2 and Appendix A). As a matter of fact, addressing the Seifert fibration has often been mentioned as a major improvement to achieve further insight at least into the mathematical side of the subject \cite{KeBiPe,LaSuMo}.

The knotted solid torus emerging from the Seifert (not Hopf) fibration shares a number of interesting features (Section 2) which comply in a natural way with experimental settings where light beams are  framed: the electric and magnetic fields oscillate in perpendicular directions ($\mathbf{B} \cdot \mathbf{E} =0$) inside the tubular neighbourhood $\mathcal{N}$ of a knot $\mathcal{K}$  endowed with a framing. We have indeed shown that the ``knottiness"  of $\mathcal{N}$ can be properly formalised by choosing a specific framing that we have called \textit{preferred framing}. Namely, one has to choose the vertical convention on the self--linking number as in Eq.\eqref{framing3}, and  adopt the Rolfsen basis convention as in Eq.\eqref{framing9}. The latter requirement is due  to the fact that the self--homeomorphisms of a (topological) 2-torus $T^2$ can be extended to the  knotted solid torus  if and only if any meridian curve on $T^2$ is mapped to a meridian in $\mathcal{N}$ bounding a disk. As a consequence of this choice --which can be regarded as a sort of gauge-fixing-- the topological invariant $\mathsf{n}$ (the number of positive or Dehn twists) can be used also to attain a simple presentation of  Seifert surfaces, \textit{cf.} Eq.\eqref{preferred3}. 
On the field--theoretical side, the choice of the preferred framing has allowed us to express the ``Seifert--Ra\~{n}ada" optical fields through an original extension of the Seifert fibration map, \textit{cf.} Eq. \eqref{area4}.\\
In Section 3 we have discussed the algebraic--topological framework of 
 ``braided open books",  an approach not considered so far in the context of braided optical fields. The issue here is the concept of ``fibrability" of simple-strand braids. By resorting to the seminal work of Rudolph \cite{Rud1,Rud2,Rud3} (see also \cite{Bode}), we have stressed that braids with braid index $\leq 3$, as well as all the classes of homogeneous braids, are indeed fibrable (Section 3.3). We argue that these achievements 
comply with current (and maybe forthcoming) experimental setups as far as the braiding of optical fields can be experimentally controlled. The ``topological constraints" in the title of this paper refer essentially to these findings.\\

In Section 4 we have investigated  the possibility of resorting to more effective topological invariants associated to framed knots, braided open books and Seifert surfaces. The simplest (numerical) ones, namely the braid index and the genus, are defined as \textit{minimal} among the collections  of strands representing the same knot, and among the genera of the surfaces derived from different knot diagrams, respectively. The search for these quantities falls in the class of NP-complete problems in geometric group theory or combinatorial topology and, as such, has a purely theoretical character \cite{BiBr}. On the other hand, the Alexander--Conway polynomial invariant (Section 4.3 and Appendix B) might  be interesting also in our context since it can be evaluated combinatorially on knot diagrams and encodes a blending of  algebraic and  topological properties. In particular, since twice its degree is the lower bound for the genus of the Seifert surface it might be used to parametrise in some weak sense the admissible Seifert surfaces. Moreover, looking at different planar diagrams of a same knot (or closed braid) the resulting different twist and  writhe numbers, together with the multiple forms of the Alexander polynomials, might be arranged as to set up some cryptographic protocols.\\
For the sake of completeness we have included in Appendix B also the definitions of other polynomial invariants of knots, highlighting the fact that they were recognised as quantum expectation values of Wilson--loop operators in $SU(2)$ or $SU(N)$ Topological Quantum Field Theories of Chern--Simons type. Topological quantities associated to electromagnetic knots, being rooted in  an $U(1)$ gauge theory,  are in contrast linking numbers or self--linking numbers that can be evaluated analytically through Gauss-type linking integrals.
We argue that taking into account (quantum)  angular momentum states of  paraxial light beams would open the possibility to introduce in this context suitable (unitary) representations of the braid group, and then to address richer algebraic structures. As a further remark, it is worth mentioning a few results found by one of the author and collaborators. A spin network quantum simulator based on the recoupling theory of $SU(2)$ quantum angular momenta has been proposed, highlighting its role as a discretised counterpart of topological quantum computing \cite{MaRa1}, as well as a generalised quantum circuit model \cite{MaRa2}. From this scheme, classes of automaton models able to evaluate efficiently $SU(2)$--coloured
polynomial invariants of knots (generalisations of Jones's invariant) have been derived
{\it cf.} \cite{GaMaRa}. Since one of the most intriguing applications of knotted optical configurations is the search for secure cryptographic protocols like the one proposed in 
\cite{LaDeFe}, we argue that more ``structured" braided light beams might be processed
within such a spin network computational scheme.\\  
As an aside remark about  quantisation --based now on the 
Riemann-- Silberstein formulation of the electromagnetic field related to Ra\~{n}ada's theory of electromagnetic knots--  we mention  the paper \cite{MaRaRa} where
the Lie algebra of the group $SU(1,1)$
--whose group manifold is an open solid torus--, instead of the customary Weyl algebra $h(1)$, has been proposed as a natural unifying frame for characterising boson systems. This perspective  applied to $U(1)$ knotted optical fields would deserve further investigations.

%\vfill
%\newpage

\section*{Appendix A. The Hopf fibration: an  overview}
For a rigorous treatment of the subject we refer the reader for instance to 
\cite{Mon}. However, there can be found quite accurate descriptions and visualisation in the physical literature on optical fibred configurations, see {\it e.g.} \cite{KeBiPe}. In this paper we  describe in some detail how homogeneous  coordinates can be introduced  as to  comply with the complex parametrisations 
adopted to deal with the Hopfion solution quoted in the Introduction.\\

\begin{figure}[H]
	\includegraphics[width=7cm]{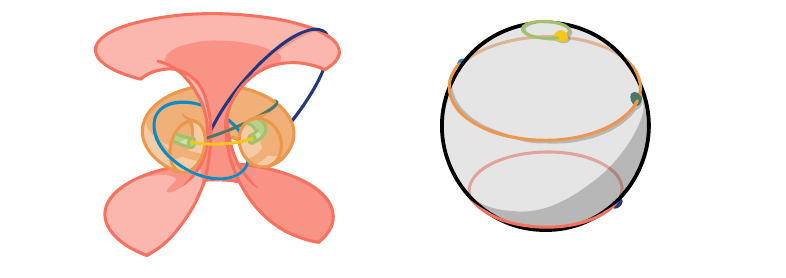}
	\centering
	\caption{Nested tori and their relative projection to circles on $S^2$.}
	\label{HOPF}
\end{figure}

\noindent The Hopf fibration is a fibre bundle structure involving spheres of suitable dimensions 
\begin{equation}
	S^1\,\hookrightarrow \,S^3 \,\xrightarrow{h}S^2\,.
	\label{hopf1}
\end{equation}
The fibre  is the circle $S^1$ embedded in the 3-sphere  --the total space-- and  the \textit{Hopf map}, $h:S^3\rightarrow S^2$, is the projection on the base space.
Being a fibre bundle, the Hopf fibration is locally indistinguishable (diffeomorphically) from the product space of an open set $A$ contained in the base space and the fibre space, $A \times S^1$
(local trivialisation property of the bundle structure). 
The inverse  $h^{-1}$ of the Hopf map takes a point $\in S^2$ to a circle $\subset S^3$, and pairs of points are mapped to a linked pair
of circles: the linking number is called the Hopf index. 
The total space of the Hopf fibration can thus be visualised as a  collection of disjoint copies of $S^1$ that fills the entire space and  has the property that any two such circles are linked to one another, generating the surface of a torus.  Owing to the trivialisation property mentioned above, $S^3$ is foliated with two families of nested (solid) tori, the \textit{Clifford tori}, as sketched in Fig. \ref{HOPF}.
The degenerate torus in the very core of the nested ensemble (indistinguishable from an $S^1$) would be projected to the North pole of the base space $S^2$, while the degenerate torus at infinity (which is indistinguishable from a vertical, straight line through the centre of the nested tori) would be projected to the South pole of $S^2$.
These are called the ``exceptional fibres" of the Hopf fibration, while the other ones are named
``ordinary fibres".\\

\noindent A first step in introducing the aforementioned coordinates consists in complexifying the manifolds, viewed as embedded in spaces of one dimension higher. Denoting $\cong$  diffeomorphisms, and $\iota_{2,3}$ regular
inclusions, the ordered triple of manifolds in \eqref{hopf1} reduce to
\begin{align}
S^1 &\cong \, \mathbb{C}^*  \nonumber \\
S^3 \cong   \mathbb{R}^3 \cup \{\infty \} & \xrightarrow{\iota_3}\; \mathbb{R}^4 \,\cong \mathbb{C}^2  \nonumber \\
S^2 \cong \mathbb{C} \cup \{\infty \}
& \xrightarrow{\iota_2}\; \mathbb{R}^3 \,\cong \mathbb{C} \times \mathbb{R}\,, 
\label{hopf2}
\end{align}
where $\mathbb{C}^*$ represents unimodular complex numbers.
A further step calls into play stereographic projections of spheres 
(in dimension 2 and 3) onto flat spaces of the same dimensions 
\begin{align}
\mathsf{sp}_3 \,:S^3 & \,\longrightarrow \, \mathbb{R}^3\,\cong \,\mathbb{C} \times \,  \mathbb{R} \label{hopf3} \\
\mathsf{sp}_2 \,:S^2 & \,\longrightarrow \, \mathbb{R}^2\,\cong \,
\mathbb{C} \,.
\label{hopf4}
\end{align} 
Of course stereographic (or homogeneous) coordinates on the spheres are local and it is necessary to specify the centre of the projection (North or South pole). In standard Euclidean coordinates a unit $n$-dimensional sphere $S^n$  is defined as $\{(x_1,x_2,\dots , x_{n+1})$
$\in \mathbb{R}^{n+1}\, | \sum_{k=1}^{n+1} x_k^2 =1 \}$. Projecting from the North pole $(0,0,\dots, 1)$ and denoting $\mathbf{x}$ the 
$n$-ple $(x_1,x_2,\dots, x_n)$, the projection map  
\begin{equation}
\mathsf{sp}_n \,: \,S^n \, \longrightarrow \, \mathbb{R}^n
\label{hopf5}
\end{equation}
is given by
\begin{equation}
\mathbf{u}  \,:= \mathsf{sp}_n\,(\mathbf{x})\,=\,
\frac{\mathbf{x}}{1-x_{n+1}}
\label{hopf6}
\end{equation}
and is a diffeomorphism from the open set $S^n \setminus \{\text{North}\}$
to $\mathbb{R}^n$. The inverse map 
$(\mathsf{sp}_n)^{-1}\,: \mathbb{R}^{n} \rightarrow S^n \subset \mathbb{R}^{n+1}$ reads explicitly
\begin{equation}
(\,\mathsf{sp}_n)^{-1}\, (\mathbf{x})  \,= \Big(
\frac{2 \mathbf{u}}{1+ u^2}\, , \, \frac{1- u^2}{1+ u^2}
\Big)\,,
\label{hopf7}
\end{equation}
with $u^2=\sum_{i=1}^n \,u_i^2$.\\

\noindent \textbf{Coordinatisation of the 2-sphere.}
Given $(x_1,x_2,x_3)$ such that $\sum_{k=1}^{3} x_k^2 =1$, and denoting $\mathsf{z}=x_1 + i x_2$,
the equation of the unit 2-sphere is obviously $|\mathsf{z}|^2 + x_3^2=1$. The complexified homogeneous coordinate is given by
\begin{equation}
S^2 \cong \mathbb{C} \cup \{\infty \}\; \xrightarrow{\mathsf{sp}_2} \;\mathbb{C} \, \ni \,
X_1 + i X_2 \equiv Z := \Big( \frac{\text{Re}\, \mathsf{z}}{1-x_3},\, \frac{\text{Im} \,\mathsf{z}}{1-x_3} 
\Big)\,,
\label{hopf8}
\end{equation}
and the coordinatisation of the inverse map $\mathbb{C} \,\rightarrow \, \mathbb{C} \cup \{\infty\} \subset \mathbb{C} \times
\mathbb{R}$ reads
\begin{equation}
(\mathsf{z},x_3) \,=\, \Big(\frac{2\,\text{Re}\, Z}{1+ \,|Z|^2}\,,\,
\frac{2\,\text{Im}\, Z}{1+ \,|Z|^2}\,,\,\frac{Z \bar{Z} -1}{1+ \,|Z|^2}
\Big)\,.
\label{hopf9}
\end{equation}
\noindent In applications to optical fields it is customary to employ cylindrical/polar coordinates: in these coordinates,  $(r,\theta, z) \in S^2 \subset \mathbb{R}^3$
and $(R, \Theta) \in \mathbb{R}^2$,  the expressions in Eqs.\eqref{hopf8} and
\eqref{hopf9} become respectively
\begin{align}
(R, \Theta) \,=& \Big(\frac{r}{1-z}\,, \theta \Big) \nonumber \\
(r, \theta, z) \,=& \Big( \frac{2R}{1+R^2}\,, \Theta \,, \frac{R^2-1}{1+R^2}\Big)\,.
\label{hopf10}
\end{align}
\noindent \textbf{Coordinatisation of the 3-sphere.}
Given $(x_1,x_2,x_3,x_4)$ such that $\sum_{k=1}^{4}$ $x_k^2 =1$, and denoting $\mathsf{z}_1 =x_1 + i x_2$, $\mathsf{z}_2 =x_3 + i x_4$,
the equation of the unit 3-sphere is  $|\mathsf{z}_1|^2 + |\mathsf{z}_2|^2 =1$. Then
\begin{align}
S^3 \cong \mathbb{R}^3 \cup \{\infty \}\; \xrightarrow{\mathsf{sp}_3} &\;\mathbb{C} \times \mathbb{R}\, \subset \mathbb{C}^2 \ni \,
(X_1 + i X_2, X_3 + i X_4) \equiv (Z_1,Z_2) \nonumber \\
\; &:= \Big( \frac{ \mathsf{z}_1}{1-x_4},\, \frac{x_3}{1-x_4} 
\Big)\,,
\label{hopf11}
\end{align}
and the inverse application reads
\begin{equation}
(\mathsf{z}_1,\mathsf{z}_2) \,=\, 
\frac{1}{1+ |Z_1|^2  + X_3^2} \big(\,
2\,\text{Re}\, Z_1, 2\,\text{Im}\, Z_1\,, 2\,\text{Re}\, Z_2\,, |Z_1|^2 + X_3^2 -1\,\big)\,.
\label{hopf12}
\end{equation}
%%%%%%%%%PARTE NUOVA
\noindent \textbf{The Hopf map and generalisations.} The action of the Hopf map
$h: S^3 \rightarrow S^2$ using the above coordinatisations is given by 
\begin{equation}
h\,(\mathsf{z}_1,\mathsf{z}_2) \,=\, \big(2 \mathsf{z}_1\mathsf{z}_2,\,|\mathsf{z}_1|^2 - 
|\mathsf{z}_2|^2\, \big)
\,,\label{hopf13}
\end{equation}
where $[\,h(\mathsf{z}_1,\mathsf{z}_2)\,]^2\,=\,[\, |\mathsf{z}_1|^2 +
|\mathsf{z}_2|^2\,]^2$ gives the unit 2-sphere $x_1^2+x_2^2+x_3^2 = 1$
with the obvious identifications $\mathsf{z}_1 = \mathsf{z}$ and $x_4 =0$ in Eq.
\eqref{hopf11}. The features of the Hopf bundle \eqref{hopf1} can be recovered by noting that: {\bf i)} if there exists a pair $(\mathsf{w}_1,\mathsf{w}_2)$ such that 
$h(\mathsf{w}_1,\mathsf{w}_2) \,=\,
h(\mathsf{z}_1,\mathsf{z}_2)$ then necessarily $\mathsf{w}_1 = \zeta \mathsf{z}_1$
and $\mathsf{w}_2= \zeta \mathsf{z}_2$ with $\zeta \in \mathbb{C}, |\zeta|^2 =1$;
{\bf ii)} if two points $\in S^3$ differ by such a $\zeta$, then 
$h(\zeta \mathsf{z}_1, \zeta \mathsf{z}_2)$ $=h(\mathsf{z}_1,\mathsf{z}_2)$, and the inverse image of any point $m \in S^2$ is a circle.\\

Recall that
the explicit parametrisation needed to unveil the decomposition of $S^3$ in nested Clifford tori would require the choice of Hopf coordinates, not the stereographic ones. However, it is more useful for our purpose to address this decomposition by looking at the following (complex) action of $S^1$ on $S^3$, namely 
\begin{align}
S^1 \times S^3 \, \rightarrow & \:\;\;\;\;\;\; S^3 \nonumber \\
(t; \mathsf{z}_1,\mathsf{z}_2) \,\mapsto & \, \big( e^{it}\,\mathsf{z}_1,\,e^{it}\,\mathsf{z}_2 \big) \, ,
\label{hopf14}
\end{align}
where the curves parametrised by $t \in [0,1]$ warp  around and fill the Clifford tori: this is actually a simple way of looking at the topological content  of the 
Hopfion solution in the context of optical fields.\\
However, the action of Eq. \eqref{hopf14} can be generalised according to \cite{Mon}
\begin{align}
S^1 \times S^3 \, \rightarrow & \:\;\;\;\;\;\; S^3 \nonumber \\
(t; \mathsf{z}_1,\mathsf{z}_2) \,\mapsto & \, \big( e^{i \alpha t}\,\mathsf{z}_1,\,e^{i \beta t}\,\mathsf{z}_2 \big) \, ,
\label{hopf15}
\end{align}
and, if $\alpha, \beta$ are coprime integers, the ordinary fibres are torus knots (see Section 4.1)
warping around $T^2 \cong S^1 \times S^1$ representing the intersection of the two families of Clifford tori: these configurations underlie  the torus knots solutions addressed in the recent physical literature \cite{KeBiPe,LaSuMo,ArTiTr}.

%\vfill
%\newpage

\section*{Appendix B. A glimpse to topological knot theory}
\begin{figure}[H]
	\includegraphics[height=7cm]{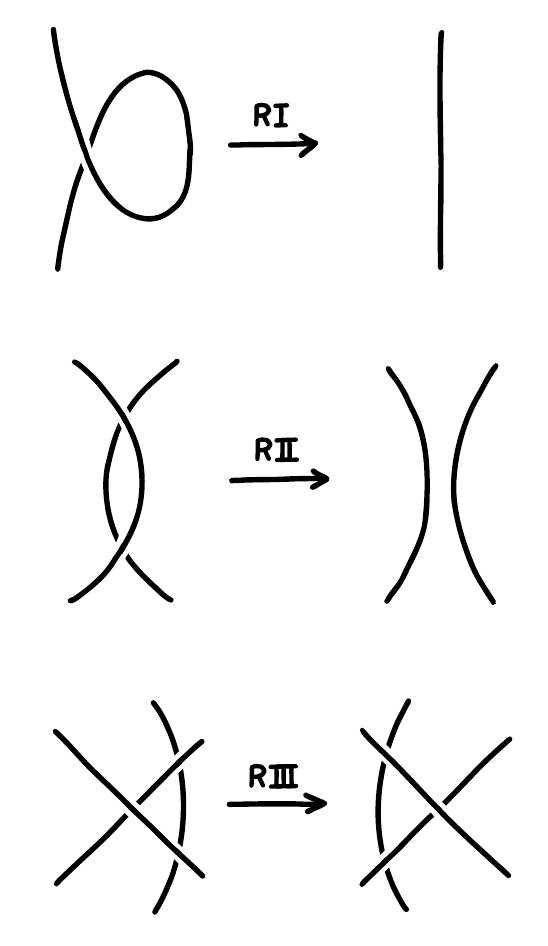}
	\centering
	\caption{The three Reidemeister moves acting on portions of a knot (link) diagram.}
	\label{Reid}
\end{figure}

A knot $\mathcal{K}$ is an embedding  of the circle $S^1$ into the Euclidean 3-space $\mathbb{R}^3$ or into the unit 3-sphere $S^3 \subset \mathbb{R}^4$ ($S^3$ can be viewed also as the compactification of 
$\mathbb{R}^3$, $\mathbb{R}^3$ 
$\cup$ $\{\infty\}$\,). A link $\mathcal{L}$ is an embedding of a finite collection of    non--intersecting circles that may be linked and knotted together, $\{L_i\} = \mathcal{L},\, i= 1,2,\dots,N$. 
Knots addressed in this paper are oriented and ``prime", namely cannot be decomposed into the connected sum of two non--trivial knots: these are the knots depicted in Knot Tables
\cite{Knots}, listed on the basis of the number of crossings. A \textbf{diagram} associated to a knot is the projection of the knot onto a plane: it is obviously not unique, but we assume that the projection has only a finite number of double points (over- and under-crossings) and no multiple points.
The existence of such  projection planes actually encodes the definition of ``tame" knots.  We agree to use the same symbol $\mathcal{K}$ for the knot and for one among its planar diagram, the meaning being specified in the text.\\
Knots are grouped into equivalence classes defined up to \textit{ambient isotopy},  continuous deformations (homeomorphism) of the knot strand in the ambient space. 
It is a classic result that two knots $\mathcal{K}$, $\mathcal{K}'$ are ambient--isotopic if and only if the associated diagrams  are related by a finite sequence of the three \textbf{Reidemeister moves} depicted in Fig.\ref{Reid}.
Throughout this paper we have used ``topological invariance"
as a synonym of ``invariance under ambient isotopy", in particular in connection with the quantities discussed in Section 4. However, we have encountered in Section 2.2 the writhe number of a diagram, $Wr (\mathcal{K})$, defined as the signed sum over the crossing points of $\pm 1$, according to the convention of Fig.\ref{incroci}. Clearly this number does not change under Reidemeister moves II and III, but does change under  move I: quantities of this type
are invariants under \textit{regular isotopy}. This remark discloses the relevance  of  the formula in  Eq.\eqref{framing2} for the self--linking number of a band, as well as the convenience of the choice of the vertical framing done in  Eq.\eqref{framing3}. \\
Note  that all the mathematical notions and results discussed in this paper are still valid for piecewise--linear knots (braids), similar to  the configurations that occur in experiments.

\vspace{10pt}

\begin{figure}[ht]
	\includegraphics[width=7cm]{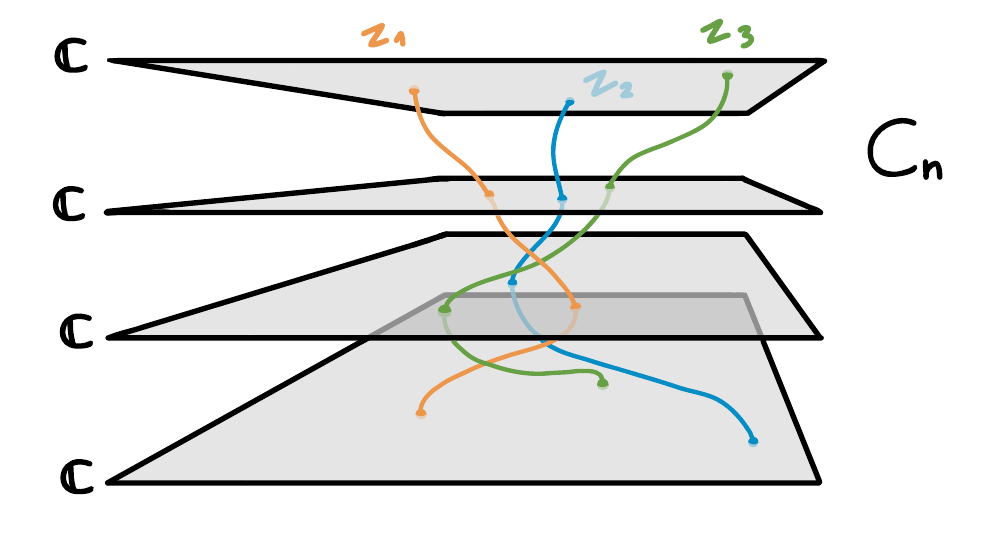} 
	\centering
	\caption{Intersecting paths representing the evolution of $n$ points in the complex plane generating a braid whose closure gives the knot curve.}
	\label{path}
\end{figure}

\noindent {\bf Parametrisation of knot curves and configuration space}\\
The explicit expressions  of  a few knot curves  (for the trefoil, cinquefoil,  figure-8 k and  lemniscate knots) have been used in \cite{KeBiPe,LaSuMo,BoDeFo} but of course it is difficult to finding out   complex analytic functions describing other knot curves. The fact that in principle one can associate a polynomial to  curves representing the evolution of ``punctures" in the complex plane,
{\it cf.} Fig. \eqref{path}, relies on a different perspective on the braid group. The basic object would be the \textit{configuration space} $\mathcal{C}_n$ on $n$ points in the  $n$-dimensional complex space,  
$\mathcal{C}_n :=$  $\{(z_1, z_2, \dots, z_n) \;\in \mathbb{C}^n :\, z_i \neq z_j \, \text{for} \,i \neq j \} $, \textit{cf.} Fig.\ref{path}. It can be shown that, up to the action of the symmetric group, the 
homotopy group of this space is B$_n$. We refer the reader to \cite{BiBr} for details, just noting that the resulting  polynomial functions of degree $n$  in a complex variable $Z$
would read $Z^n + a_1 Z^{n-1} + \dots + a_{n-1} Z +a_n$, where the coefficients $(a_1, \dots a_n)$
are polynomials in the $\{(z_1, z_2, \dots, z_n)\}$. This kind of polynomials are not to be confused with the topological polynomial invariants of knots discussed below.

\vspace{10pt}

\noindent {\bf Properties of the Alexander invariant}\\
\textbf{i)} Owing to the ambiguity in selecting the Seifert surface of a given knot (see Section 4.2),
$\Delta_{\mathcal{K}}(t)$ is only defined up to multiplication by $t^{\pm N}$, with $N \geq 1$. This remark can be related to the known fact that twice the degree of the Alexander polynomial is a lower bound for the genus $g$ of the surface. Consider for instance the right--handed trefoil knot (depicted in Fig.\ref{TREF} below): the standard expression gives $\Delta$ (right trefoil) $= t^2-t +1$, and its genus is $1$, as can be easily checked from the formula for torus knots given in Section 4.2 recalling that this knot 
has $(p,q)=$ $(2,3)$. However, multiplying by $t^{-1}$, one get also $\Delta$ (right trefoil) $= t-1+t^{-1}$, an expression often preferred since the powers match for positive and negative. \\
\noindent \textbf{ii)} As we have seen in Section 4.1, torus knots are always fibrable since they are associated to (positive or negative) homogeneous braids, a property  established in Section 3.2. More generally, according to the result found in \cite{Neu}, the Alexander polynomial of any \textit{fibred knot} is always monic.
This seems to provide a criterion for detecting ``fibrability'' but the converse statement is not always true: it holds for alternating knots and knots up to 10 crossings,
see \cite{Knots}.\\
\noindent \textbf{iii)} The Alexander--Conway polynomial is quite effective in distinguishing inequivalent knots: it classifies completely knot diagrams  with $\leq 8$ crossings but, as said previously, cannot detect mirror images. A knot is ``chiral" if it is not equivalent to is mirror image, ``achiral" if it is. The trefoil is chiral (the two forms are shown in Fig.\ref{TREF}) and thus  $\Delta$ (right trefoil) $= \Delta$ (left trefoil).
The same holds true for all of the torus knots, for which  $(p,q)$ represents the right--handed and $(p,-q)$ the left--handed configuration. $\blacksquare$
\begin{figure}[ht]
	\includegraphics[width=8cm]{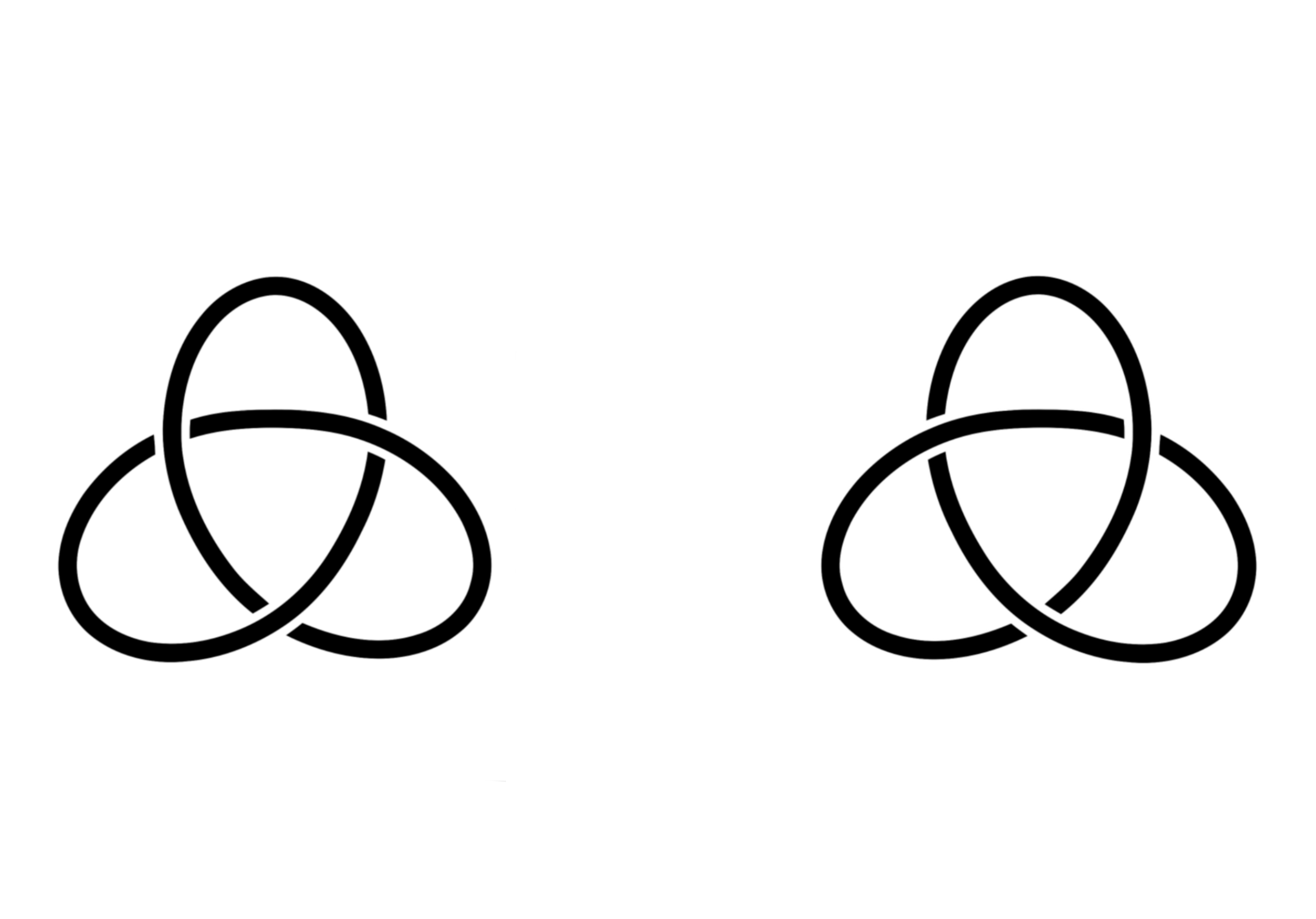}
	\centering
	\caption{The two inequivalent trefoil knots: left--handed (left) and right--handed (right).}
	\label{TREF}
\end{figure}

\vspace{10pt}
The homological construction of the Alexander polynomial is essentially topological  and attempts to investigate other approaches of this kind in the search for invariants of knots were not quite successful. The crucial ingredient to go ahead has been recognised to be rooted in braid theory (Section 3.1),
and more specifically in looking for invariant quantities associated to (matrix) representation theory of braid groups $B_n$. Indeed, Birman proved that the Alexander polynomial can be obtained from the Bureau representation, the only known at that time \cite{Bir}.

\vspace{10pt}

\noindent {\bf Other polynomial invariants}\\
The {\bf Jones polynomial} was discovered by resorting to representations of $B_n$
in von Newmann algebras \cite{Jon85}, and later on was reformulated in several ways, 
from representations in Hecke algebras \cite{Jon87,BiBr} up to representations in the universal enveloping algebra of the Lie algebra $sl(2)$ \cite{ReTu} and \cite{Wit,Gua} (in the context of quantum field theory).
In terms of skein relations on knot diagrams, the Jones polynomial $V_{\mathcal{K}} \in \mathbb{Z}[t^{1/2}, t^{-1/2}]$ is recovered from
\begin{equation}
t^{-1}V_{\mathcal{K}_+} - tV_{\mathcal{K}_-} =\left( \sqrt{t} - \frac{1}{\sqrt{t}}\right)  
V_{\mathcal{K}_0} \,,
\label{jones1}
\end{equation}
with $V_{\bigcirc} = 1$.
All knots (and their mirror images) with $\leq 9$ crossings have distinct Jones invariants. Torus knots share a simple general formula for their $V(t)$
and of course are good examples of knots with the same Alexander's and different Jones's invariants: for the two trefoils one has
\begin{equation}
V (\text{right trefoil}) \,=\, t^3 + t - t^4 \,; \;\;\;V (\text{left trefoil}) \,=\,
t^{-1} + t^{-3} -t^{-4}\,.
\label{jones2}
\end{equation}
However, there are knots with higher numbers of crossings which have both the same Jones' and Alexander's polynomials. The inequivalent 11-crossing knots named after Conway
(11n34 in Thistlewaite notation) and Kinoshita--Terasaka (11n42) \cite{Knots} share this property, and in particular their Alexander invariant is the same as the trivial knot. (These knots can be distinguished through their genus, 3 and 2 respectively, but, as discussed at length in Section 4.2, there is no efficient algorithm to calculate the genus in general.)\\
The  two invariants   discussed so far
possess a common generalisation given by  the 2-variable  
{\bf HOMFLY polynomial} $P_{\mathcal{K}}\in \mathbb{Z}[\alpha, w] $, whose name stands for the initials of those who contributed to its discovery \cite{HOMFLY}. It can be calculated using the skein relation
\begin{equation}
\alpha \,P_{\mathcal{K}_+} - \alpha^{-1}  \,P_{\mathcal{K}_-} =w \,P_{\mathcal{K}_0} \,,
\label{homfly}
\end{equation} 
and  admits also an interpretation in the context of $SU(N)$ quantised gauge theories of topological type, see \textit{e.g.} \cite{Gua}. The Alexander--Conway polynomial is recovered
by choosing $\alpha =1$, and Jones' by setting $\alpha = t^{-1}$ and 
$w= (\sqrt{t} -1) / \sqrt{t}$.\\
It is worth mentioning that Rudolph in \cite{Rud3} proposed  a 2-variable polynomial derived from a suitable splitting of the dimension of the first homology group of the Seifert surfaces in braided open books. However, we are not able to reconstruct such an expression within the ``standard'' frameworks for knots invariants discussed so far.

%\end{document}


\vfill
\newpage
  
 
\begin{thebibliography}{99}

\bibitem{IrBo} Irvine W T M and Bouwmeester 2008
	Linked and knotted beams of light {\it Nat. Phys.} {\bf 4} 716

	\bibitem{Irv} Irvine W T M 2010 Linked and knotted beams of light, conservation of helicity and the flow of null electromagnetic fields {\it J. Phys. A: Math. Theor.} {\bf 43} 385203
	arXiv: 1110.5408
	
\bibitem{KeBiPe} Kedia H, Bialyniki--Birula I, Peralta--Salas D and  Irvine W T M 2013
Tying knots in light fields {\it Phys. Rev. Lett.} {\bf 111} 150404 arXiv:1302.0342


\bibitem{BoDeFo} Bode B, Dennis M R, Foster D and King R P 2016 Knotted fields and explicit fibrations for lemniscate knots {\it Proc R. Soc. A} {\bf 473} 20160829 arXiv:1611.02563


\bibitem{ArTiTr}  Array\`as M, Tiemblo and Trueba J L 2023
	The quest of null electromagnetic knots from Seifert fibration
	{\it Chaos, Solitons Fract.} {\bf 166} 113002 
	
	
	\bibitem{LaSuMo} 
Larocque H, Sugic D, Mortimer D, Taylor A J, Fickler R, Boyd R W, Dennis M R and Karimi E 2018 Reconstructing the topology of optical polarization knots {\it Nature Phys.} {\bf 14} 1079 https://doi.org/10.1038/s41567-018-0229-2 
	
	\bibitem{SuDrOt} Sugic D, Droop R, Otte E, Ehrmanntraut D, Nori F, Roustekoski J, Denz C and Dennis M R 2021 Particle-like topologies in light 
{\it Nat. Commun.} {\bf 12} 6785  


\bibitem{LaDeFe}  Larocque H, D' Errico A, Ferrer--Garcia M F, Carmi A, Cohen E and Karimi E 2020 Optical framed knots as information carriers {\it Nat. Commun.} {\bf 11} 5119 


\bibitem{Ran89} Ra\~nada A F  1989 A topological theory of the electromagnetic field 
{\it Lett. Math. Phys.} {\bf 18} 97
	
\bibitem{Ran92} Ra\~nada A F  1992 Topological electromagnetism {\it J. Phys. A: Math. Gen.}  {\bf 25} 1621
	
\bibitem{RaTr95} Ra\~nada A F and  Trueba J L 1995 Electromagnetic knots
{\it Phys. Lett. A} {\bf 202} 337

\bibitem{RaTr06} Ra\~nada A F and  Trueba J L 2006 Topological quantization of the magnetic flux
{\it Found. Phys.} {\bf 36} 427 


\bibitem{ArBoTr} Array\`as M, Bouwmeester D and Trueba J L 2017 Knots in Electromagnetism 
{\it Phys. Rep.} {\bf 667} 1

\bibitem{Whi} Whitehead J H C 1947 An expression of Hopf's invariant as an integral
{\it Proc. Natl. Acad. Sci. USA} {\bf 33} 117

\bibitem{Ale1} Alexander J W 1923 A lemma on a system of knotted curves
 {\it Proc. Natl. Acad. Sci. USA} {\bf 9} 93

	\bibitem{Rud1} Rudolph L 1987 Isolated critical points of maps from $\mathbb{R}^4$ to $\mathbb{R}^2$ and a natural splitting of the Milnor number of a classical fibred link. Part I: Basic theory;  examples {\it Comment. Math. Helvetici} {\bf 62} 630 
arXiv: math/0203032v2 [math.GT]
	
	\bibitem{Rud2} Rudolph L 1986 Isolated critical points of maps from $\mathbb{R}^4$ to $\mathbb{R}^2$ and a natural splitting of the Milnor number of a classical fibred link. Part II 
{\it Geometry and Topology: Manifolds, Varieties and Knots}  (Boca Raton FL: CRC Press)  p 251
	
	\bibitem{Rud3} Rudolph L 1988 Mutually braided open books and new invariants of fibered Links {\it Contemp. Math.} {\bf 78} {\it Proc. of the AMS-IMS-SIAM Joint 
Res. Conf. on Artin's Braid Group 1986} ed J Birman and A Libgober {Providence RI: Amer. Math. Soc.} p 657
	
	\bibitem{Har} Harer J 1982 How to construct all fibered knots and links {\it Topology} {\bf 21} 263
	
	
	\bibitem{Sta} Stallings J R 1978  Constructions of fibred knots and links {\it Proc. Symp.  Pure Math. Amer. Math. Soc.} {\bf 32} 55
	
	\bibitem{Tesi} Sanna N 2022 {\it Optical fields and topological structures: geometric--topological characterisation of optical framed knots} Master Thesis (Universiy of Milano Bicocca-I, available on request)


\bibitem{SeTh} Seifert H and Threlfall W 1980 {\it A Textbook of Topology}
	 (New York: Academic Press)
	 
	 
	 \bibitem{Mon} Montesinos J M 1987 Classical Tessellations and Three--Manifold 
(Berlin Heidelberg: Springer--Verlag)

\bibitem{BaGr}  Baader S and Graf C 2016 Fibred links in $S^3$ 
{\it Expo. Math.} \textbf{34} 423



\bibitem{Car} C\u alug\u areanu G 1961 Sur le classes d'isotopie des noeuds tridimensionels et leur invariants {\it Czech. Math. J.} {\bf 11} 588


\bibitem{Rol} Rolfsen D 1976 {\it Knots and Links} (Berkeley CA: Publish or Perish, Inc.)
	

\bibitem{Lic} Lickorish W B R 1997 {\it An Introduction to Knot Theory} (New York Berlin: Springer)

\bibitem{Gua} Guadagnini E 1993 {\it The Link Invariants of the Chern--Simons Field Theory} (Berlin New York: Walter de Gruyter)	




\bibitem{Pul} Pulemotov A 2011 {\it Prescribed Ricci curvature on a solid torus}
arXiv:1106.2605 [math.DG]


\bibitem{BiBr} Birman J S and Brendle T E 2004 {\it Braids: a Survey} {Handbook of knot theory} W Menasco and M Thistletwaite ed (Amsterdam: Elsevier) 2006
 arXiv:0409205v2 [math.GT]

\bibitem{Knots} {\it The Knot Atlas} 
$\text{https:}//\text{katlas.org}/$


\bibitem{Bode} Bode B 2023 Braided open book decompositions in $S^3$ {\it Rev. Mat. Iberoam}
{\bf 39} 2187
arXiv:2111.05187v2

\bibitem{Mor} Morton H R 1985 Exchangeable braids from low--dimensional manifolds
{\it LMS Lecture Notes} {\bf 95} ed R A Fenn (Cambridge UK: Cambridge Univ. Press) p 86

\bibitem{Yam} Yamada S 1987 The minimum number of Seifert circles equals the braid index of a link {\it Invent. Math.}  {\bf 89} 347

\bibitem{BiMe}  Birman J S and Menasco W W 2008 A note on closed 3--braids 
{\it Commun. Contemp. Math.} {\bf 10} 1033

\bibitem{Ale2} Alexander J W 1928 Topological invariants of knots and links
{\it Trans. Amer. Math. Soc.} {\bf 30} 275

\bibitem{Jon85}Jones V F R 1985 A polynomial invariant for knots via von Neumann algebras 
{\it Bull. Amer. Math. Soc.} {\bf 12} 103


\bibitem{Wit}
Witten E 1989 Quantum field theory and the Jones polynomial
{\it Commun. Math. Phys.} {\bf 121}  351

\bibitem{Kau} Kauffman L H 2001 {\it Knots and Physics} Series on Knots and Everything 
Vol. 1 $3^{rd}$ Edition (Singapore: World Scientific) 

\bibitem{Con} Conway J H 1970 An enumeration ok knots and links and some of their algebraic properties {\it Computational Problems in Abstract Algebra}
p 329 (New York: Pergamon Press)
	

\bibitem{MaRa1}
Marzuoli A and Rasetti M 2005 Computing spin networks
{\it Ann. Phys.} {\bf 318} 345 arXiv:quant-ph/0410105

\bibitem{MaRa2}
Marzuoli A and Rasetti M 2017 Spin network quantum circuits 
{\it Int. J. Circ. Theor. Appl.} {\bf 2017} 
DOI: 10.1002/cta.2346


\bibitem{GaMaRa}
Garnerone S,  Marzuoli A and Rasetti M 2009 Efficient quantum processing
of three-manifold topological invariants 
{\it Adv. Theor. Math. Phys.}  {\bf 13} 1601 arXiv:quant-ph/0703037
 
\bibitem{MaRaRa}
Marzuoli A, Raffa F A and Rasetti M 2014 Where do bosons actually belong?
{\it J. Phys. A: Math. Theor.} {\bf 47} 275202 arXiv:1406.2908

\bibitem{Neu}	Neuwirth L 1960 The algebraic determination of the genus of knots
{\it Amer, J. Math} {\bf 82} 791
	
\bibitem{Bir} Birman J S 1975 {\it Braids, Links , and Mapping Class Groups} 
(Princeton NJ: Princeton University Press)


\bibitem{Jon87}Jones V F R 1987 Hecke algebra representations of braid group and link polynomials
{\it Ann. of Math.} {\bf 126} 335

\bibitem{ReTu} Reshetikhin N and Turaev V G 1991 Invariants of 3-manifolds via link polynomials and quantum groups {\it Invent. Math.} {\bf 103} 547 
doi.org/10.1007/BF01239527

\bibitem{HOMFLY} Freyd P, Yetter D, Hoste J, Lickorish W B R, Millett K and Ocneanu A
1985 A new polynomial invariant of knots and links {\it Bull. Amer. Math. Soc.}
{\bf 12} 239

\end{thebibliography}
\end{document}